\documentclass[12pt]{iopart}
\usepackage{iopams}  
\usepackage{epsfig}
\usepackage{graphicx}

\begin{document}

\title[Flow in Hybrid Approaches]{Anisotropic flow in transport+hydrodynamics hybrid approaches}

\author{Hannah Petersen${}^{1,2}$\\[.4cm]}

\address{
${}^1$~Frankfurt Institute for Advanced Studies, Ruth-Moufang-Strasse 1, D-60438 Frankfurt am Main, Germany\\
${}^2$~Institut f\"ur Theoretische Physik, Goethe Universit\"at, Max-von-Laue-Strasse 1, D-60438 Frankfurt am Main, Germany\\
}

\ead{petersen@fias.uni-frankfurt.de}

\begin{abstract}
This contribution to the focus issue covers anisotropic flow in hybrid approaches. The historical development of hybrid approaches and their impact on the interpretation of flow measurements is reviewed. The major ingredients of a hybrid approach and the transition criteria between transport and hydrodynamics are discussed. The results for anisotropic flow in (event-by-event) hybrid approaches are presented. Some hybrid approaches rely on hadronic transport for the late stages for the reaction (so called afterburner) and others employ transport approaches for the early non equilibrium evolution. In addition, there are 'full' hybrid calculations where a fluid evolution is dynamically embedded in a transport simulation. After demonstrating the success of hybrid approaches at high RHIC and LHC energies, existing hybrid caluclations for collective flow observables at lower beam energies are discussed and remaining challenges outlined. 
\end{abstract}


\pacs{25.75.-q,25.75.Ag,24.10.Lx,24.10.Nz}


\maketitle

\section[Introduction]{Introduction}
\label{intro}
Collective flow is one of the earliest proposed observables to measure the properties of strongly interacting matter at high temperatures and densities created in ultra-relativistic collisions of heavy nuclei \cite{Baumgardt:1975qv,Danielewicz:1985hn,Ollitrault:1992bk}. The interaction strength of the whole system is reflected by common trends in the behaviour of particles coming from a nuclear collision. Collective behaviour of the produced particles in a heavy ion collision is manifested  for example as radial flow, a common velocity of different particle species that can be explained by a fireball with a certain temperature and collective velocity, one of the first pieces of evidence that a new state of matter, an equilibrated quark gluon plasma, is formed. 

This manuscript is focussed on anisotropic collective flow, which can be quantified by a Fourier expansion in momentum space in the azimuthal direction in the transverse plane. The most prominent of these Fourier coefficients is the second one the so called elliptic flow $v_2$ \cite{Ollitrault:1992bk}. In a non-central heavy ion collision the overlap region between the two nuclei is not spherically symmetric, but exhibits an ellipsoidal deformation. Conventionally the impact parameter direction is defined to be the x-direction which corresponds to the shorter axis of the ellipse. If all particles would be produced with an isotropic cross section independently of each other the resulting distribution in momentum space would be rotationally symmetric. But if the hot and dense nuclear matter interacts collectively the coordinate space asymmetry is translated to a momentum space asymmetry and can be measured as anisotropic flow. Therefore, collective anisotropic flow provides crucial information about the transport properties of strongly interacting matter at high temperatures and densities. Even if the particles are non-uniformly distributed in coordinate space, a rather strong interaction between the particles is needed to transfer the geometrical granularity to observable structures in momentum space, e.g. higher moments of the Fourier decomposition.  

In 2005, 5 years after the Relativistic Heavy Ion Collider (RHIC) was turned on, the announcement that the quark gluon plasma has been found and behaves like an almost perfect liquid was mainly based on the observation of jet quenching and high elliptic flow \cite{Adams:2005dq,Back:2004je,Arsene:2004fa,Adcox:2004mh}. The agreement between elliptic flow measurements and hydrodynamic predictions was key to this conclusion \cite{Kolb:2000fha,Kolb:2001qz,Huovinen:2001cy} and has been confirmed by viscous hydrodynamic calculations later on\cite{Luzum:2008cw,Song:2007ux}. 

\begin{figure}[h]
\centering
\includegraphics[width=0.5\textwidth,angle=-90]{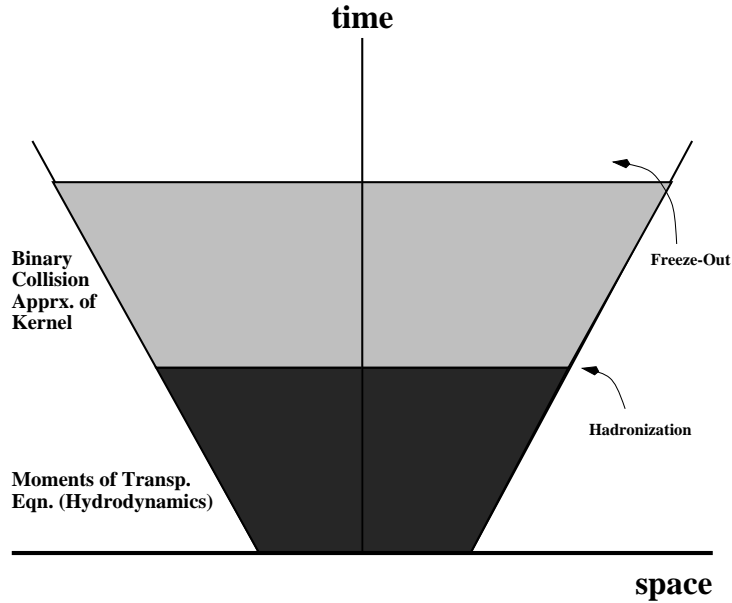}
\caption{Schematic overview of the space-time evolution of a
high-energy heavy-ion collision as assumed in hybrid approaches. Fig. taken from \cite{Bass:2000ib}.} \label{fig_sketch_hic} 
\end{figure}

Already in the early days of the success of ideal hydrodynamic calculations at RHIC, it became clear that a sudden transition from a strongly coupled system with vanishing mean free path to free streaming particles on a Cooper-Frye transition surface is not sufficient. Therefore, hybrid approaches were proposed that combine the advantages of hydrodynamics with the ones from hadronic transport \cite{Bass:2000ib, Teaney:2000cw}. A schematic overview of the space-time evolution of a heavy ion reaction within a hybrid approach is shown in Fig. \ref{fig_sketch_hic}. In the later stages of the reaction the particle abundancies are fixed first and then the kinetic freeze-out occurs. Also, different particle species decouple successively from the fireball according to their hadronic cross sections. All these effects can be accommodated in so called hybrid approaches that match a hydrodynamic evolution to a hadronic cascade. 

Within the last ~15 years hybrid approaches have been proven to be very successful in describing global features of heavy ion collisions at RHIC and the Large Hadron Collider (LHC) \cite{Hirano:2012kj} (and references therein). One can even say, that combined models applying (viscous) hydrodynamics for the hot and dense stage and hadronic transport for the later dilute stages of the reaction have been established as the 'standard model' for the dynamical description of high energy heavy ion collisions. After the realization that event-by-event fluctuations are crucial and a wealth of anisotropic flow and correlation observations by the experimental collaborations is becoming available, the opportunity for a more quantitative assessment of quark gluon plasma properties using state-of-the-art dynamical approaches is emerging \cite{Adare:2012kf,Luzum:2013yya} (and references therein). 

This article presents an overview on anisotropic flow in hybrid approaches. The scope is limited to the strict definition of hybrid approaches by selecting only results from calculations where a hydrodynamic evolution is combined with a transport approach. This restriction is necessary to keep the article focussed and means to exclude interesting work on transport approaches that exploit hydrodynamic features, but also hydrodynamic calculations that do not match to transport calculations at any point. 

The structure of the article is as follows: In Section \ref{hybrid} the basics of hybrid approaches are introduced including a description of the main features of transport and hydrodynamics and the transition criterion between both.  In Section \ref{normal_hybrid} the traditional hybrid approaches where transport is applied for the late stages of the reaction are reviewed and results for elliptic flow calculations are presented. To study the effect of initial state fluctuations one can also employ a transport approach for the early non-equilibrium stage of the reaction and create the initial state for the fluid dynamical calculation event by event. This type of hybrid approach is presented in Section \ref{fluc_ic}. The last option is to combine hydrodynamics and transport in such a way that transport approaches are used for the initial and the final stages of the reaction and hydrodynamics is sandwiched between two dynamical stages with non-equilibrium properties. These 'full' hybrid approaches are highlighted in Section \ref{full_hybrid}. The last Section \ref{flow_exc} is devoted to the study of anisotropic flow as a function of the beam energy within hybrid approaches. Finally, Section \ref{concl} contains a summary of the major points and an outlook on future challenges.

\section[Traditional Hybrids]{Hybrid Basics}
\label{hybrid}
Hybrid approaches are very successful to describe the dynamical evolution of heavy ion collisions at RHIC and LHC energies \cite{Muller:2012zq, Huovinen:2013wma}. The idea behind hybrid approaches is to combine the advantages of transport and hydrodynamics. Fluid dynamic calculations are based on conservation laws

\begin{equation}
\partial_\mu T^{\mu\nu}=0 \quad \mbox{and } \quad \partial_\mu N^\mu=0,
\end{equation}
where $T^{\mu\nu}$ is the energy-momentum tensor and $N^\mu$ is the net baryon
current. 

Local equilibration is assumed and only macroscopic information about the system is available. The main advantage of hydrodynamics is that the equation of state is an explicit input and allows for a direct connection to QCD properties as calculated on the lattice \cite{Aoki:2006we} or by phenomenological models. Nearly ideal hydrodynamics  is applicable to the hot and dense stage of a heavy ion reaction where the density is high and the mean free path is small. Recently, full (3+1) dimensional viscous hydrodynamic approaches are being developed and successfully applied to calculate observables in heavy ion collisions \cite{Schenke:2010nt,Vredevoogd:2012ui,DelZanna:2013eua,Akamatsu:2013wyk,Bozek:2011ua}. The main focus has been on the influence of shear viscosity, even though there are more transport coefficients of interest, e.g. the bulk viscosity or heat conductivity. The interplay between bulk viscosity and shear viscosity has been studied in \cite{Monnai:2009ad,Song:2009rh,Bozek:2009dw,Denicol:2009am,Denicol:2010tr,Roy:2011pk,Dusling:2011fd} indicating that there might be non-trivial effects that are important for a more quantitative assessment of transport properties of the quark gluon plasma \cite{Bozek:2012fw,Noronha-Hostler:2013gga,Noronha-Hostler:2014dqa}. 

In the early and the late stages of the reaction, the assumption of local equilibration cannot be justified, therefore non-equilibrium transport approaches are better suited. Transport calculations are based on the Boltzmann equation with a collision kernel that is usually truncated at the 2-body level

\begin{equation}
\label{boltzmann}
p^\mu \cdot \partial_\mu f_i(x^\nu, p^\nu) = \mathcal{C}_i \quad .
\end{equation}

A microscopic description of the whole phase-space distribution is provided which is based on hadronic or partonic degrees of freedom and their respective interaction cross-sections. Even though there are attempts to describe a phase transition within transport approaches, the microscopic mechanism for hadronization is poorly understood so far. 

A combination of hydrodynamics with transport approaches has become the standard state-of-the-art to cover the whole evolution of heavy ion reaction at high energies in a realistic way. Hydrodynamics allows for an assessment of macroscopic features e.g. the equation of state and transport properties like the shear and bulk viscosity while hadronic transport is valuable to model the decoupling process including a separated chemical and kinetic freeze-out. 

One of the most crucial points in a hybrid approach is the question when to switch from the macroscopic hydrodynamic to the microscopic transport description. Ideally there is a region in which the two calculations are equivalent and a change in the transition criterion does not affect the results for observables. In general, it is important that the equation of state on both sides of the transition surface is the same, i.e. a hadron gas with the same degrees of freedom as the transport approach of choice propagates. 

In practice, the transition from the fluid dynamic regime to the microscopic description is treated as follows. The quark gluon plasma is converted to hadronic degrees of freedom via the equation of state that is an input to the hydrodynamic description. The transition hypersurface for particlization (the conversion from fluid to particles) needs to be determined according to the Cooper-Frye formula
\begin{equation}
E \frac{dN}{dp^3}= \frac {g} {(2\pi)^3}  \int_\sigma d\sigma_\mu p^\mu f(x,p)
\end{equation} 
where $f(x,p)$ is the grand-canonical boosted particle distribution function in each hypersurface element, with normal $d\sigma_\mu$, and the total momentum distribution is an integral over the entire surface $\sigma$.  Here, $g$ is the degeneracy of the particle in question. There are different techniques how to construct a proper hypersurface in 4 dimensions, one particular implementation has been described in \cite{Huovinen:2012is} and the corresponding CORNELIUS code is publicly available on the OSCAR archive \footnote{Fortran and C++ subroutines in 3D and 4D are available at https://karman.physics.purdue.edu/OSCAR/}. 

On this hypersurface particles are sampled using Monte Carlo methods. The resulting phase space distribution of hadrons is propagated further in a hadron transport approach. In viscous hydrodynamics the deviation from the equilibrium distribution function needs to be taken into account. A common approach is to parametrize the $\delta f$ term proportional to $p^2$, the so called democratic ansatz, but in principle the species dependent varaiations need to be determined by kinetic theory \cite{Denicol:2012cn}. Another issue in the sampling process are the negative contributions, that are neglected in most hybrid approaches. Depending on the relative orientation of flow velocities and hypersurface vectors $d\sigma_\mu$ a fraction of the particles flies back into the dense region that is described by fluid dynamics. Without dynamical coupling between the fluid dynamic and the microscopic regime, it is hard to take these contributions into account in a consistent fashion. Recently, an algorithmic idea has been proposed in \cite{Pratt:2014vja}. Previous studies have shown that the negative contributions to the particle spectra at midrapdity are rather small (5\% or even smaller), but they might increase at forward rapidities and when event by event fluctuations are included \cite{Huovinen:2012is,Song:2010aq}.    

An intuitive criterion for the switching between fluid dynamics and microscopic transport is that the scattering rate reaches a critical value where the mean free path is too much increased to apply hydrodynamics. In \cite{Holopainen:2012id} it has been shown, that such a dynamical criterion gives very similar results for bulk observables like the commonly applied temperature criterion. For heavy ion collisions at lower beam energies where the net baryo chemical potential is finite, an energy density criterion might be more appropriate. A systematic investigation of transition criteria where the local equilibrum assumption breaks down across different beam energies and systems sizes would be desirable.

\begin{figure}[h]
\centering
\includegraphics[width=0.5\textwidth]{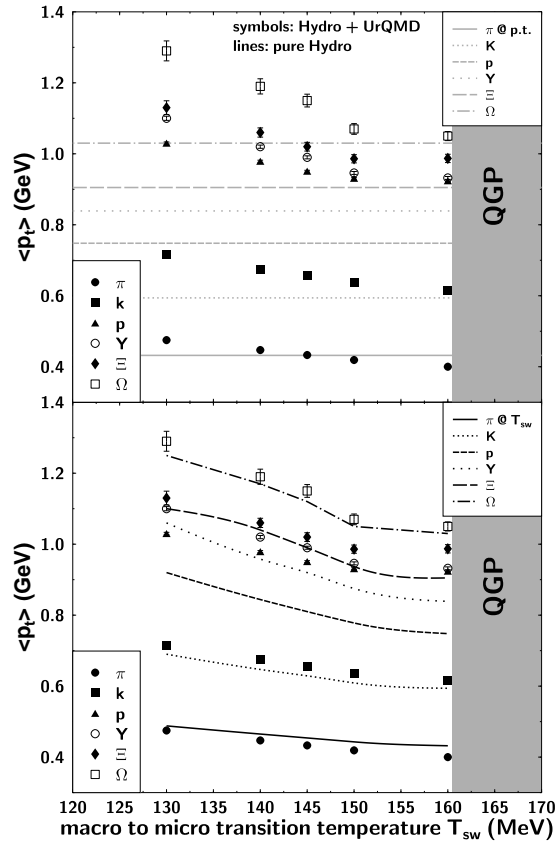}
\caption{Mean transverse momentum $\langle p_T \rangle$ 
of various hadron species at freeze-out (symbols) vs.\ the macro to micro
transition temperature $T_{sw}$. The horizontal lines in the upper frame show
the respective $\langle p_T \rangle$ values right after hadronization. The lines in the lower frame show the
$\langle p_T \rangle$ emerging from ideal flow down to $T=T_{sw}$.
This figure is for central Au+Au collisions at the highest RHIC energy. Fig. taken from \cite{Bass:2000ib}.} \label{fig_tempswitch_meanpt} 
\end{figure}

The first actual hybrid approach has been descibed in \cite{Bass:2000ib}. Figure \ref{fig_tempswitch_meanpt} shows the final mean transverse momentum for different hadron species as a function of the switching temperature between the fluid dynamic and the microscopic description. The mean transverse momentum gives an indication how much radial flow is developed during the hadronic rescattering stage of the collision. In the upper plot, the final values (symbols) are compared to the ones from the fluid dynamic evolution at a temperature of 160 MeV \footnote{This temperature was the critical hadronization temperature within the first order equation of state in this calculation at that time. Nowadays, an equation of state with a cross-over is favored for high energy nuclear collisions with a somewhat lower inflection point in temperature.}. Comparing the symbols with the grey lines shows that there is only a mass dependent increase, if the fluid evolution is run to lower temperatures. The lines in the lower figure indicate the radial flow that was generated by the hydrodynamic evolution for several hadron species. It is clearly visible that the transverse flow of protons and $\Lambda$'s is significantly increased during the non-equilibrium evolution, whereas for $\Omega$ and $\Xi$ baryons the effect is negligible. The hadronic cross-sections for the multi-strange particles are too small to change their radial flow. Pions on the other hand undergo a lot of interactions and are affected by a large amount of resonance decays, such that the final result does not differ much from the value directly at the transition.  

The general outcome from these early studies, that the multi-strange particles are frozen out at the transition surface and the protons suffer the major effect of the hadronic rescattering have been confirmed in many of the subsequently developed hybrid approaches (see following Sections). In addition the matching to a kinetic transport approach with individual particle degrees of freedom allows to generate output that is very similar to the one measured in experiment. Experiments detect a finite number of hadrons in the detector and especially for more differential correlation and fluctuation observables it has been proven to be crucial to apply the same analysis technique in the theory calculation as it has been applied for the experimental measurement \cite{Luzum:2013yya}.

\section[traditional_hybrid]{Traditional Hybrid Approaches}
\label{normal_hybrid}

\begin{figure}[h]
\includegraphics[width=0.5\textwidth]{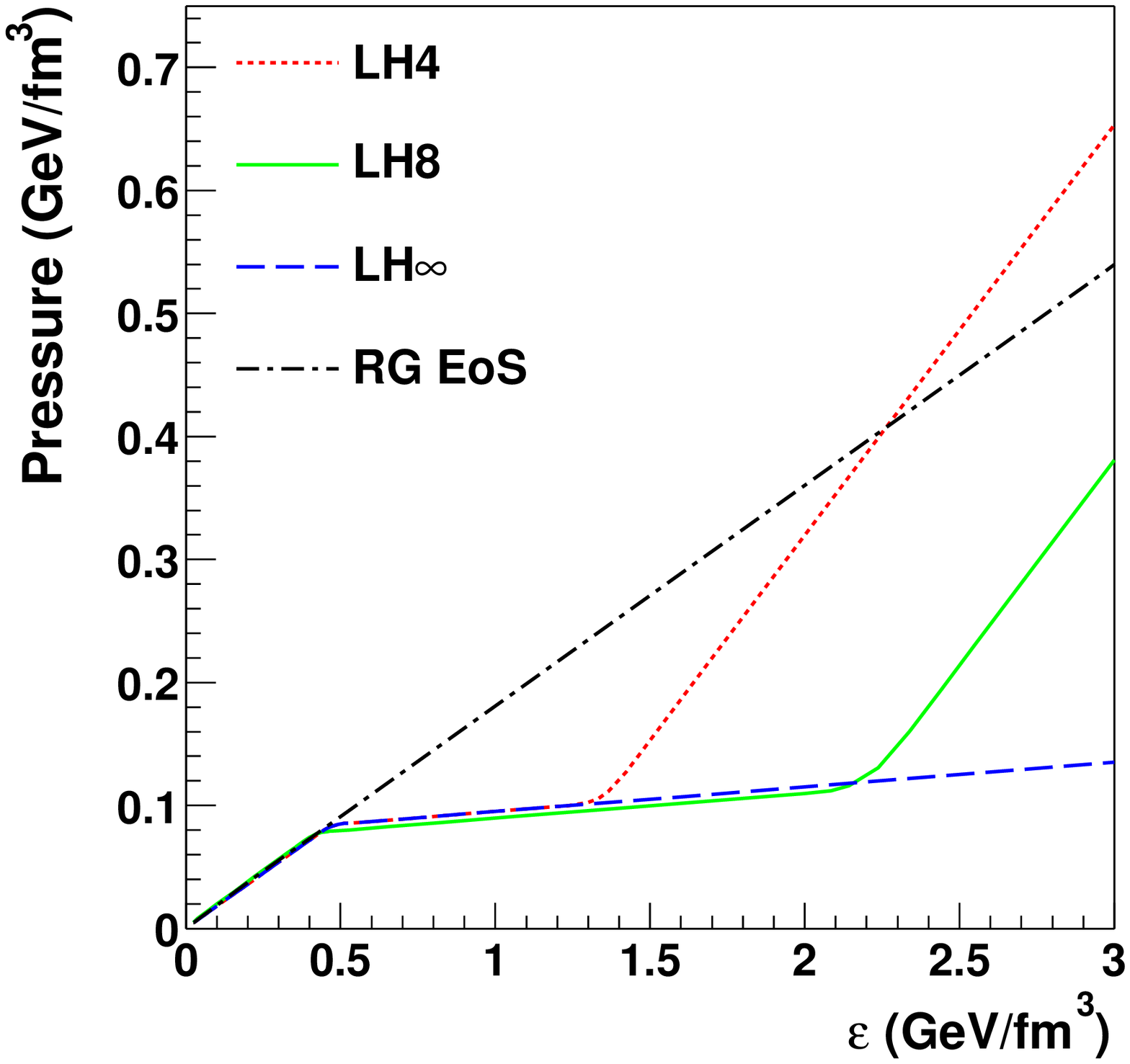}
\includegraphics[width=0.5\textwidth]{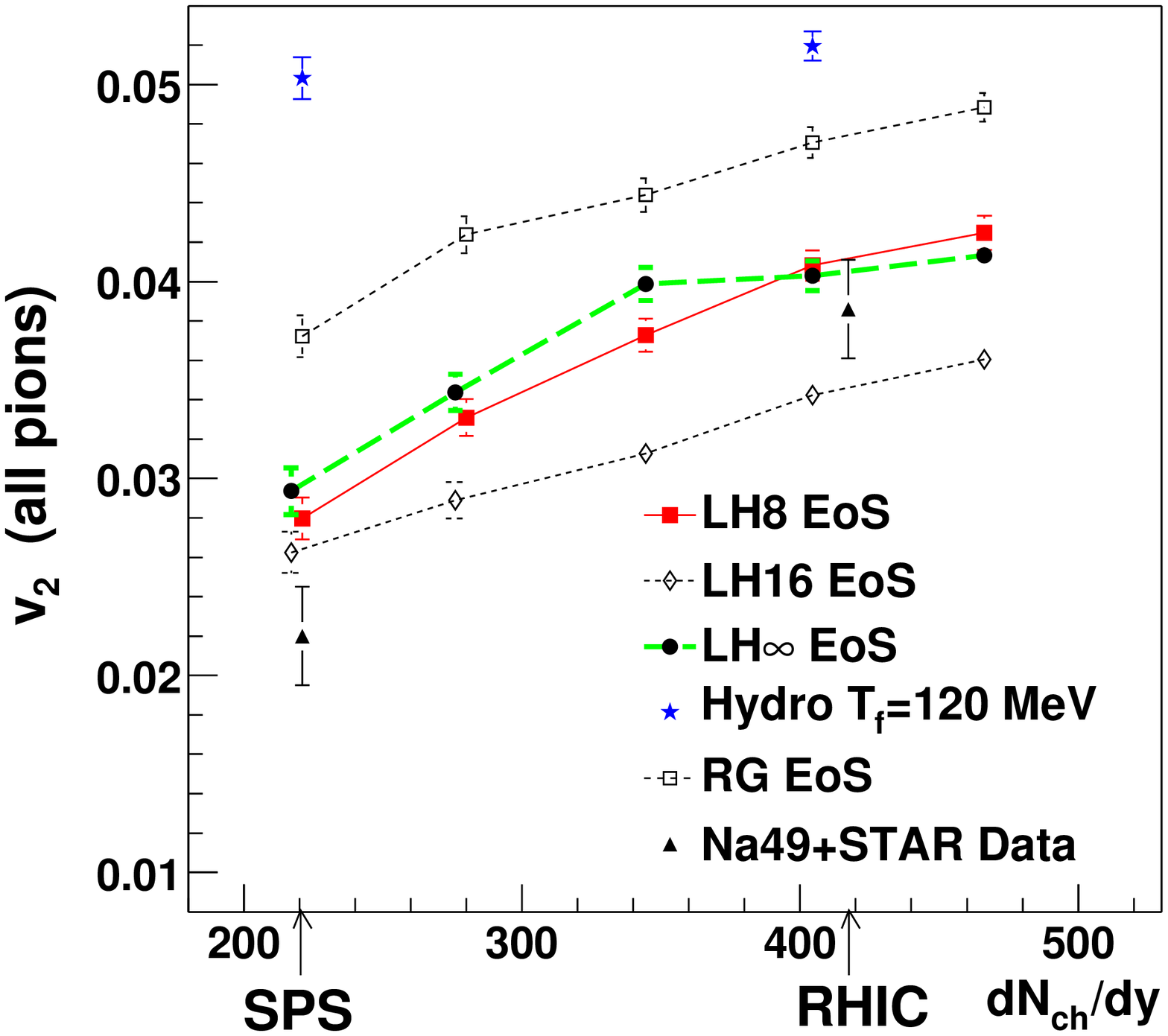}
\caption{Left: The pressure versus the energy density for different equations of state (EoS). EoSs with Latent Heats 0.4\,GeV/fm$^{3}$, 0.8\,GeV/fm$^{3}$,... are labeled as LH4, LH8,...etc. 
Right: The elliptic flow parameter $v_2$ as a function of the charged particle multiplicity in Pb + Pb collisions at an impact parameter of b=6\,fm. At the SPS, the NA49 $v_2$ data point is extrapolated to b=6\,fm using Fig.~3 in \cite{Bachler:1999hu}. At RHIC, the STAR $v_2$ data point is extrapolated to $N_{ch}/N_{ch}^{max}=0.545$ (b=6\,fm in Au+Au) using  Fig.~3 in 
\cite{Ackermann:2000tr}. 
Figs. taken from \cite{Teaney:2000cw}.} \label{fig_v2_teaney_shuryak} 
\end{figure}

Shortly after the pioneering studies in \cite{Bass:2000ib} that are shown above, further hybrid approaches have been developed. In \cite{Teaney:2000cw} the sensitivity of elliptic flow on the equation of state (EoS) has been investigated within a hybrid approach. This approach is based on boost invariant 2+1 dimensional ideal hydrodynamics coupled to RQMD at a switching temperature of 160 MeV. As shown in Fig. \ref{fig_v2_teaney_shuryak} (left) different equations of state have been constructed interpolating between a pure resonance gas (RG) and a first order phase transition with different amounts of latent heat up to infinity.  The LH$\infty$ case mimics non-equilibrium effects.

As can be seen in Fig. \ref{fig_v2_teaney_shuryak} (right) the effect of very different equations of state on elliptic flow as a function of charged particle multiplicity at two different beam energies (SPS and RHIC) is smaller than the difference between a pure hydrodynamic calculation compared to the hybrid setup \footnote{The comparison to data is a little unfair: For the model, $v_{2}$ is calculated using all pions in PbPb collisions. For the NA49 data, $v_{2}$  is measured using only $\pi^{-}$ in PbPb (a -3\% correction to the model). For the STAR data, $v_{2}$ is measured using charged hadrons in AuAu (a +5\% correction to the model).}. The blue stars indicate the rather high values of elliptic flow that are generated in a pure hydrodynamic calculation with a universal freeze-out temperature of 120 MeV at SPS energies. Once the hadronic rescattering stage is properly taken into account, the general trend of elliptic flow as a function of energy can be reproduced with any EoS. Already at that time it became clear that detailed conclusions about the properties of the phase transition to the quark gluon plasma from experimental data require a realistic treatment of the dynamical evolution. The extremly soft EoS gives almost the same results as LH8 because of canceling effects, less pressure is compensated by longer expansion. 

\begin{figure}[h]
\includegraphics[width=0.5\textwidth]{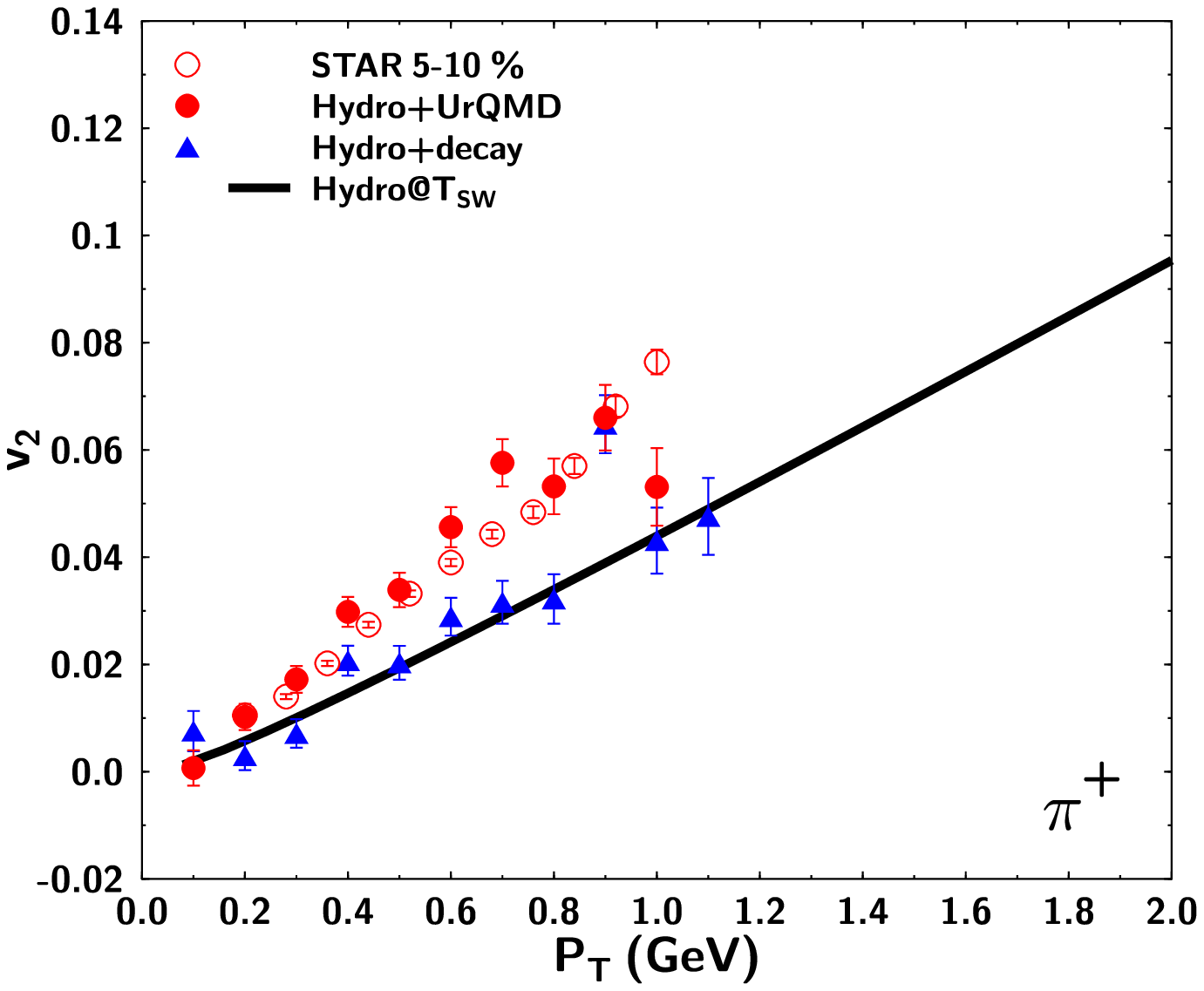}
\includegraphics[width=0.5\textwidth]{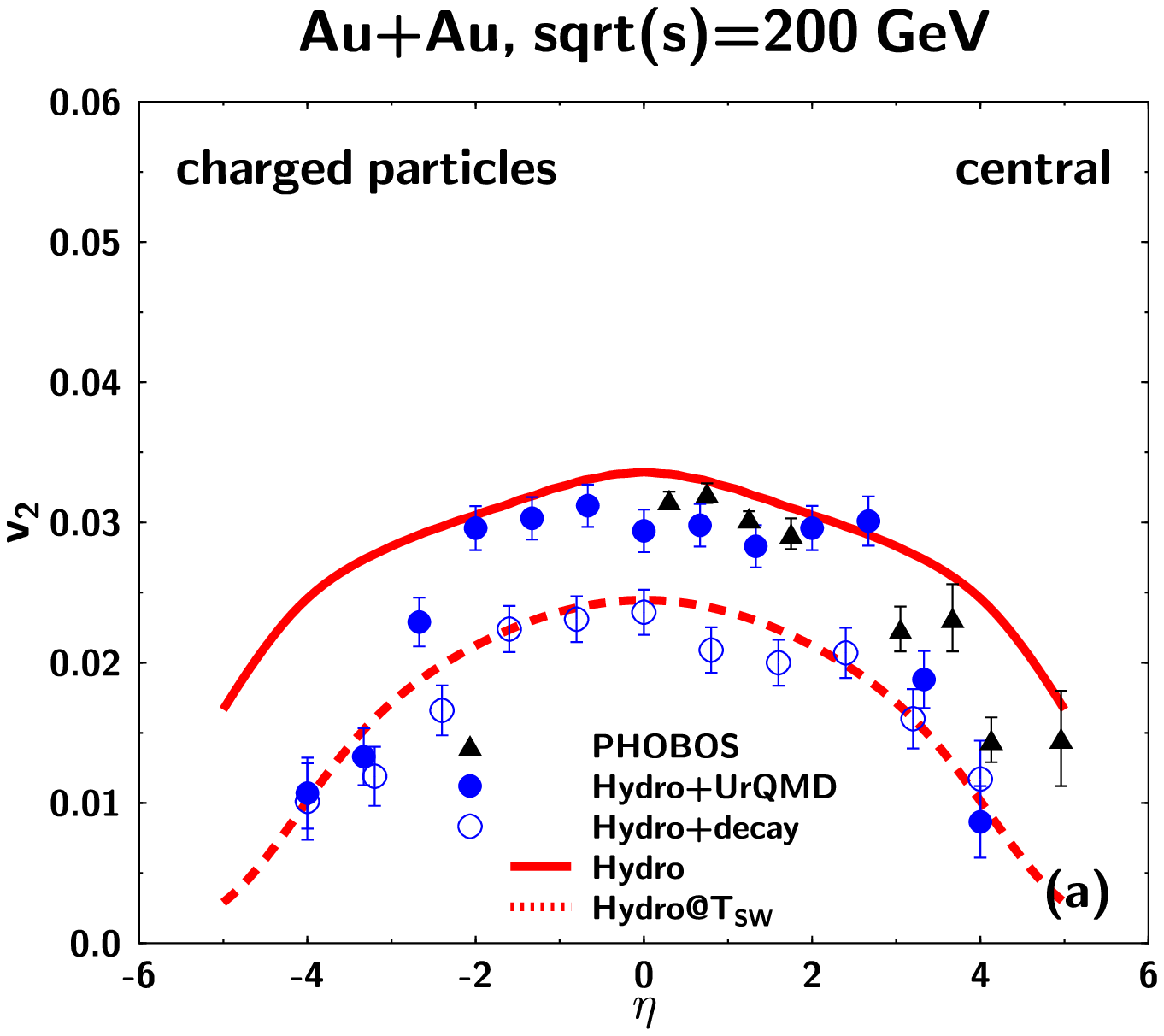}
\caption{Left: Elliptic flow as a function of $p_T$ of $\pi^+$ at centrality 5-10 \% from pure hydro at the switching temperature (solid line), hydro + decay (solid triangles) and hydro + UrQMD (solid circles). Open symbols stand for experimental data from STAR. 
Right: Elliptic flow as a function of $\eta$ of charged particles. Solid line and dashed lines stand for pure hydro calculation at the freezeout temperature (110 MeV) and the switching temperature, respectively. Experimental data by PHOBOS is shown with solid triangles. Open (Solid) circles stand for hydro + decay (hydro + UrQMD). Figs. taken from \cite{Nonaka:2006yn}.} \label{fig_v2_3dhybrid} 
\end{figure}

Subsequently, 3+1 dimensional ideal relativistic hydrodynamic approaches have been combined with hadron transport approaches. A Lagrangian hydrodynamic implementation \cite{Nonaka:2005aj,Nonaka:2006yn} has been matched to the UrQMD hadron transport approach \cite{Bass:1998ca,Bleicher:1999xi}. Again, a bag model equation of state with finite latent heat has been employed to calculate Au+Au collisions at the highest RHIC energy. In Fig. \ref{fig_v2_3dhybrid} the effects of the hadronic rescattering on elliptic flow are demonstrated. In this calculation the pseudorapidity dependence of elliptic flow has been exploited in addition to the transverse momentum dependence. In both cases, the decays alone do not influence the final anisotropic flow result, but the rescattering in the hadronic stage appears to be important to reach a reasonable agreement with the experimental data. Especially at forward rapidities the elliptic flow is reduced due to the non-equilibrium evolution while the transverse momentum dependent elliptic flow is increased. Later on, it has been shown that initial state fluctuations can have a similar effect as hadronic rescattering on these observables \cite{Andrade:2006yh,Andrade:2008xh}. 

\begin{figure}[h]
\includegraphics[width=0.5\textwidth]{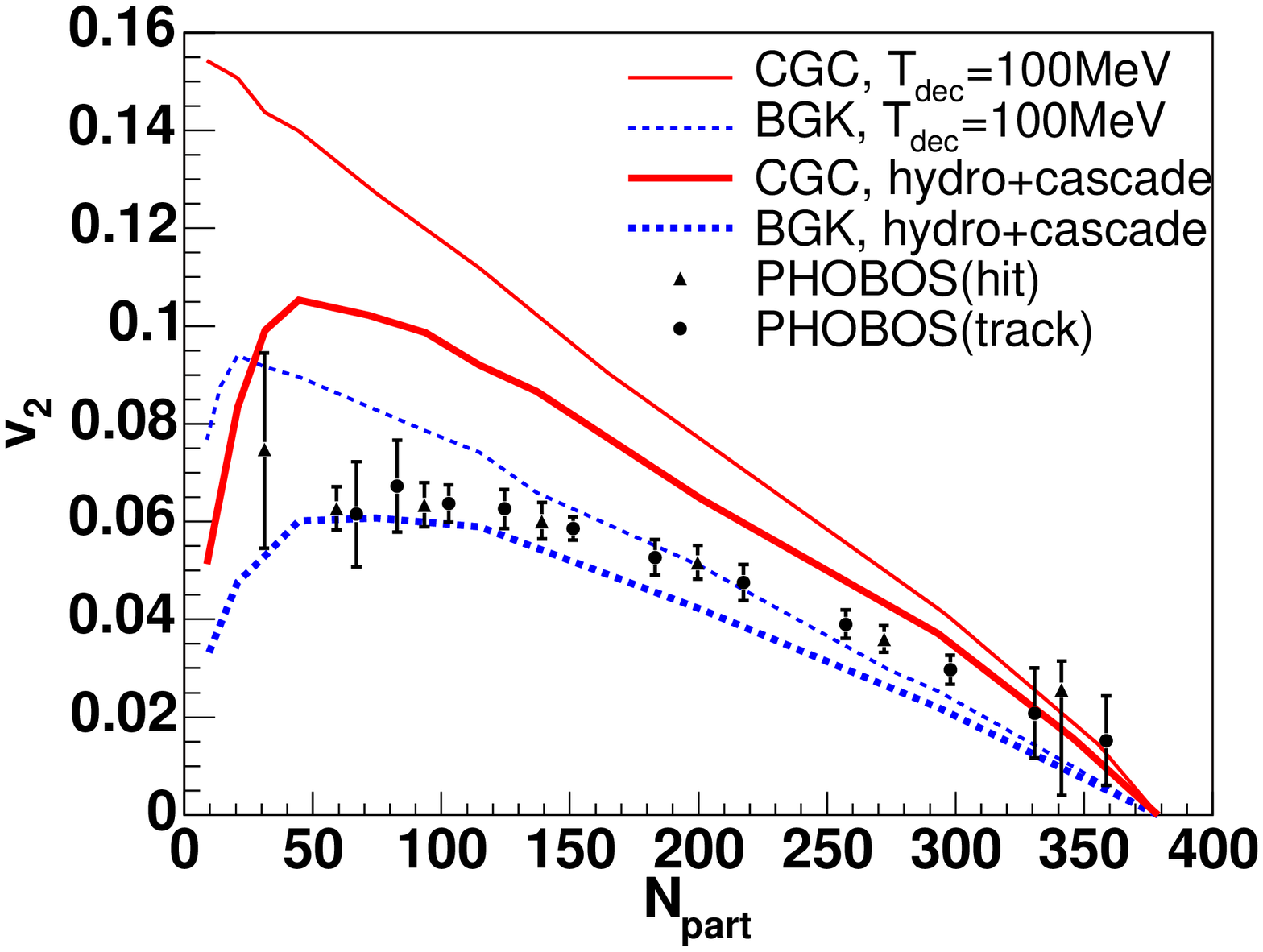}
\includegraphics[width=0.5\textwidth]{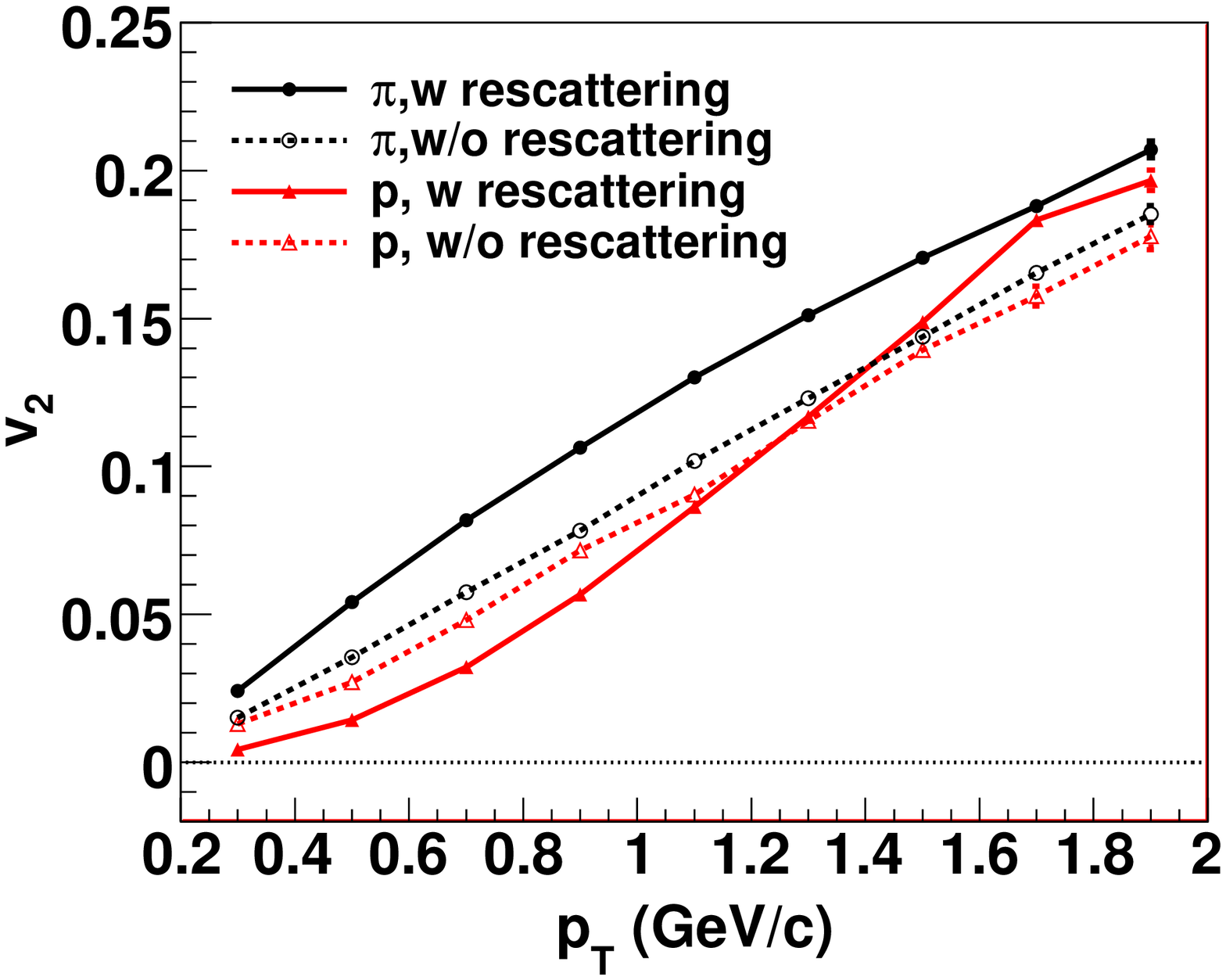}
\caption{Left: $p_T$-integrated elliptic flow for charged hadrons at midrapidity ($\mid\eta\mid < 1$) from 200\,$A$\,GeV Au+Au collisions, as a function of 
the number $N_{\rm part}$ of participating nucleons. The thin lines show the prediction from ideal fluid dynamics with a freeze-out temperature $T_{\rm dec} = 100$\,MeV, for CGC (solid red) and BGK (dashed blue) initial conditions. The thick lines (solid red for CGC and dashed blue for BGK initial conditions) show the corresponding results from the hydro+cascade hybrid model. The data are from the PHOBOS Collaboration \cite{nucl-ex/0407012}. Fig. taken from \cite{Hirano:2005xf}. Right: Transverse momentum dependence of the elliptic flow parameter
  for pions and protons. Solid (dashed) lines are with (without)  hadronic rescattering. Fig. taken from \cite{Hirano:2007ei}.} \label{fig_v2_hirano_hybrid} 
\end{figure}

Another 3+1 dimensional hybrid approach based on euclidean ideal hydrodynamics and the hadron transport approach JAM has been developed \cite{Hirano:2005xf, Hirano:2007ei}. The left hand side of Fig. \ref{fig_v2_hirano_hybrid} shows the centrality dependence of elliptic flow in gold-gold collisions at RHIC. For two different initial state parametrizations (BGK is a Glauber type and CGC is a color glass condensate type model) it is confirmed that the hadronic rescattering has an influence on the qualitative behaviour of the anisotropic flow as a function of centrality. Fig. \ref{fig_v2_hirano_hybrid} (right) demonstrates clearly the effect of the hadronic rescattering on the mass splitting of elliptic flow. Since the radial flow of protons is increased by roughly 30 \% during the hadronic stage of the reaction the elliptic flow as a function of transverse momentum gets a more pronounced shape. While the curves for protons and pions directly after the hydrodynamic evolution are almost identical the full mass splitting that has been observed at RHIC and more recently also at the LHC can only be reproduced, if the subsequent hadronic decoupling is properly taken into account \cite{Song:2013qma}. 

\begin{figure}[h]
\vspace{1.2cm}
\centering
\includegraphics[width=0.3\textwidth,angle=-90]{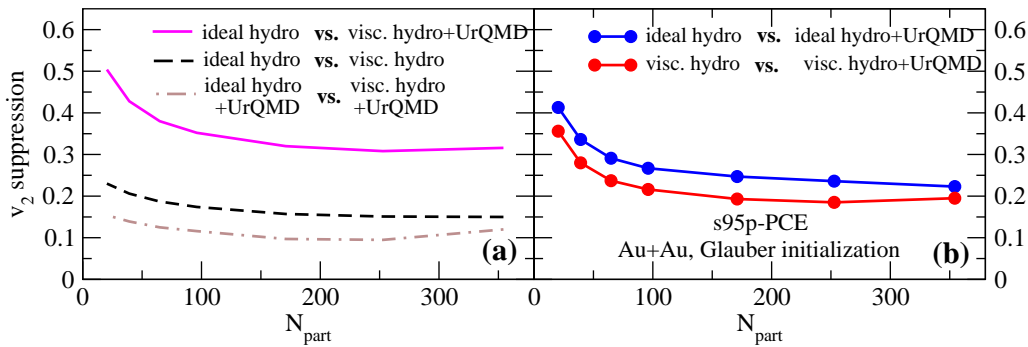}
\caption{Centrality dependence of the $p_T$-integrated elliptic flow of all hadrons. Viscous $v_2$ suppression between different scenarios as described in the text. All calculations use EOS s95p-PCE in the hydrodynamic stage \cite{Huovinen:2009yb}. Fig. taken from \cite{Song:2010aq}.} 
\label{fig_vishnu_dissipation} 
\end{figure}

Since 2008, viscous hydrodynamic calculations \cite{Luzum:2008cw,Song:2007ux} have been employed to investigate the perfect fluidity of the quark gluon plasma in more detail. The goal is to extract the transport properties of the QGP by adjusting the shear viscosity over entropy parameter in the calculation and compare to experimental data. To quantify the contribution of the final hadronic rescattering stage to the key observables like the elliptic flow a viscous hybrid approach (VISHNU) \cite{Song:2010aq,Song:2010mg,Song:2011hk,Song:2011qa,Shen:2011zc,Song:2013qma} has been developed. For this purpose 2+1 dimensional relativistic dissipative fluid dynamics has been matched to the UrQMD hadron transport approach similar to \cite{Soltz:2012rk}.   

Fig. \ref{fig_vishnu_dissipation} demonstrates that the hadronic stage is more relevant for the elliptic flow at the highest RHIC energies than the value of the viscosity in the quark gluon plasma. To quantify the suppression of integrated elliptic flow by adding a viscosity to the quark gluon plasma phase and hadronic dissipation effects the ratio
\begin{eqnarray*}
  v_2^\mathrm{supp.} = \frac{v_2^{\mathrm{A}}-v_2^{\mathrm{B}}}
                            {v_2^{\mathrm{A}}}
\end{eqnarray*}
is defined, where A and B denote different setups for the dynamical evolution. These ratios are shown as a function of centrality for different scenarios: 'ideal hydro' stands for vanishing QGP viscosity 'visc. hydro' for a minimal viscosity of $\eta/s = 0.08$ during the hydrodynamic evolution.  The decoupling temperature in the pure fluid dynamical simulations has been chosen to be $T=100$ MeV and resonance decays are taken into account. 'ideal/visc. hydro + UrQMD' denotes the full hybrid calculation with a switching temperature of 165 MeV. 

Plotting the ratios of elliptic flow developed in the different scenarios allows to quantify the relative importance of the viscosity in the different stages of the reaction. First of all, it is obvious that any viscous correction is more relevant in peripheral collisions because of the small system sizes. The pink full line in Fig. \ref{fig_vishnu_dissipation} (a) shows the maximal effect by comparing a pure ideal fluid calculation to the viscous hybrid model. The interesting observation is now that the two curves in Fig. \ref{fig_vishnu_dissipation} (b) that represent the contribution of the microscopic rescattering in the transport to ideal or viscous hydrodynamics is very similar and larger than the contribution of the minimal QGP shear viscosity (dashed lines in (a)). Therefore, it is crucial to include the hadronic transport for detailed comparisons to experimental data. 

\begin{figure}[h]
\vspace{1.2cm}
\centering
\includegraphics[width=0.7\textwidth]{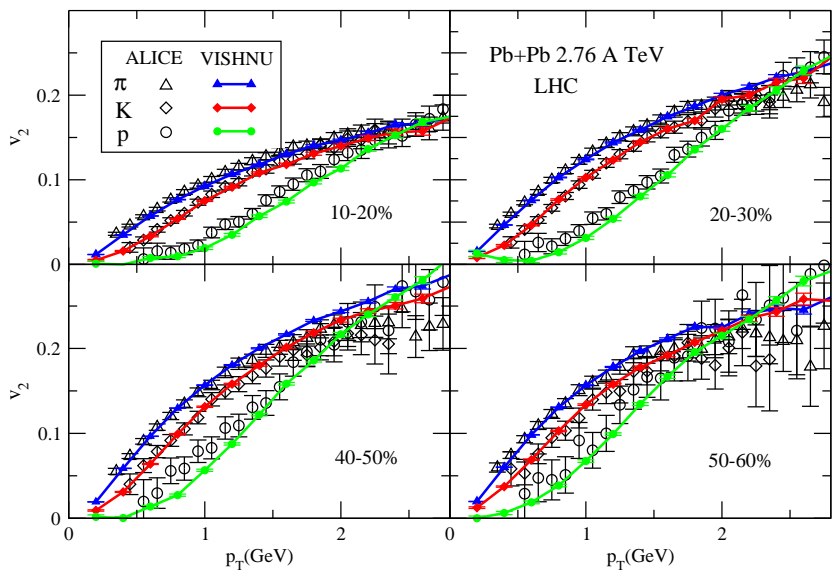}
\caption{Differential elliptic flow $v_2(p_T)$ for pions, kaons and protons in 2.76\,$A$\,TeV Pb+Pb collisions. Experimental data are from ALICE~\cite{Noferini:2012ps}. Theoretical curves are from {\tt VISHNU}. Fig. taken from \cite{Song:2013qma}.} \label{fig_vishnu_lhc_id} 
\end{figure}

Fig. \ref{fig_vishnu_lhc_id} shows the elliptic flow for identified particles as a function of transverse momentum at LHC energies for four different centralities. The viscous hybrid appraoch VISHNU is also able to describe the full mass splitting systematics at higher collision energies very well. Since this cannot be achieved in pure fluid dynamic calculations alone, this is more evidence for the need of a proper treatment of the hadronic rescattering. 

A full assessment of the contribution of the hadronic rescattering phase to bulk observables requires a 3+1 dimensional viscous hybrid approach. There is a very active development of 3+1 D viscous hydrodynamics and it is just a matter of time until full hybrid simulations will be available \cite{Ryu:2012at,vanderSchee:2013pia}. Then, the whole longitudinal expansion and structures along rapidity are accessible in addition.

\section[Fluctuating Initial Conditions]{Fluctuating Initial Conditions}
\label{fluc_ic}

To study fluctuations in the initial energy density profiles in heavy ion collisions complementary 'hybrid' approaches have been developed. In this case, the initial non-equilibrium evolution is treated with a microscopic hadron string or partonic transport approach. Even though the equilibriation process itself is out of the scope of these microscopic transport approaches, employing an off-equilibrium model for the early stage of the heavy ion reaction has some advantages compared to simplified parametrizations (Glauber etc). The fluctuations of the positions of the nucleons within the nuclei and the successive binary scatterings are taken into account. Also, the fluctuating energy deposition per collision is included as well as the straight forward possibility to match also the momentum density components of the energy momentum tensor in addition to the energy or entropy density. 

\begin{figure}[h]
\centering
\includegraphics[width=0.35\textwidth,angle=-90]{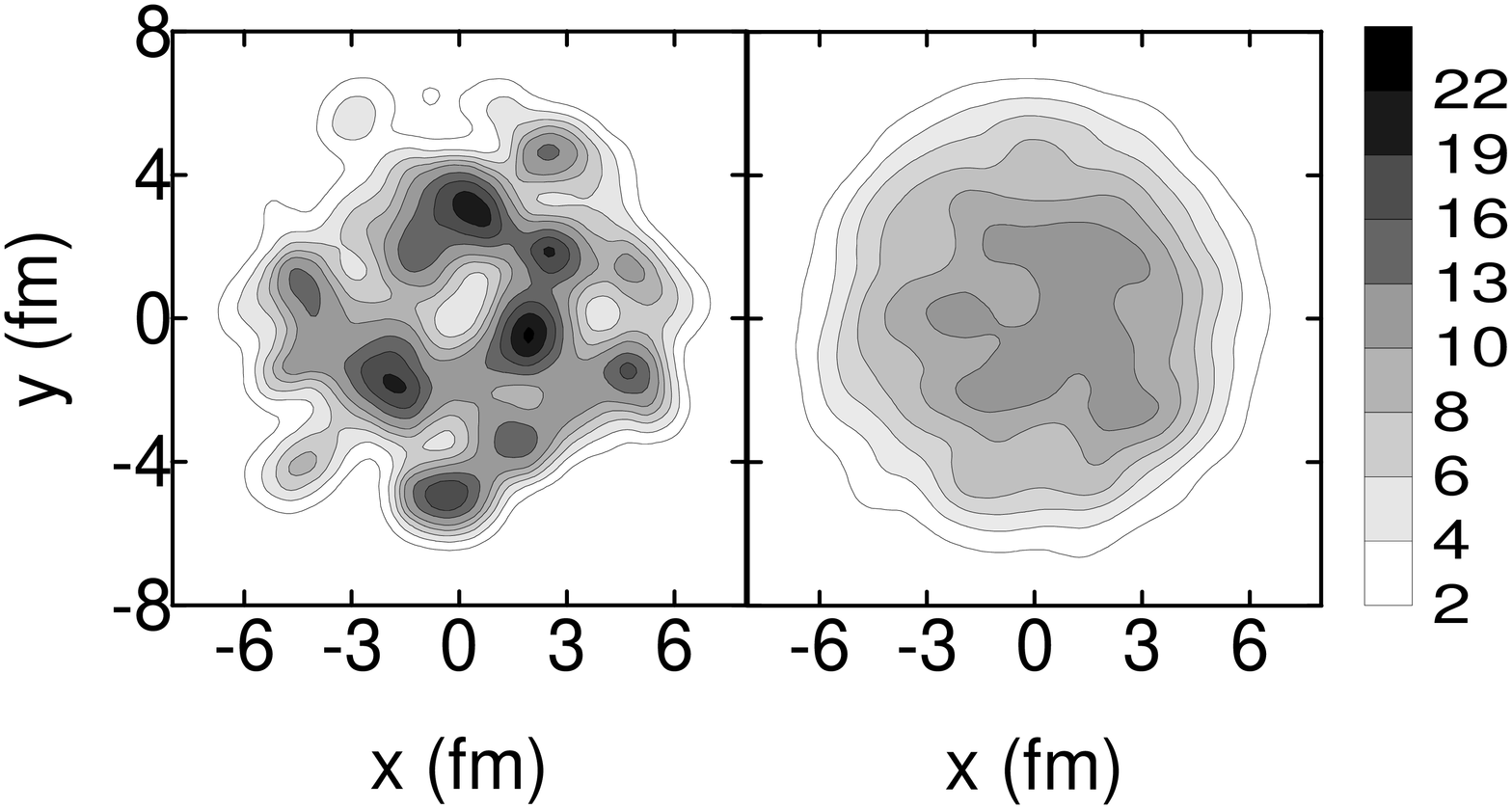}
\caption{Examples of initial conditions for 
central Au+Au collisions given by NeXus at mid-rapidity 
plane. The energy density is plotted in units of 
GeV/fm$^3$. One random event compared to average over 30 random 
events. Fig. taken from \cite{Andrade:2008xh}.} \label{fig_v2_nexspherio1} 
\end{figure}

\begin{figure}[h]
\centering
\includegraphics[width=0.5\textwidth]{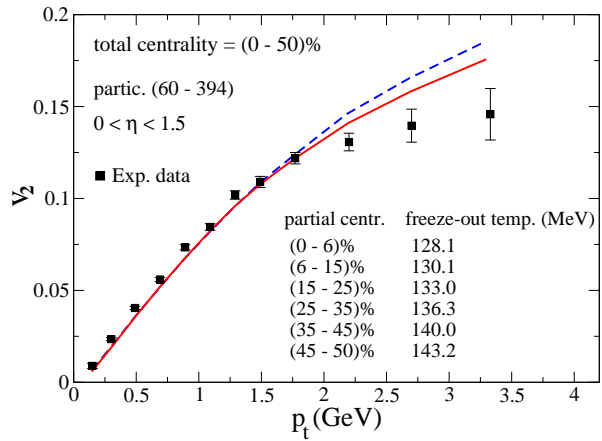}
\caption{$p_T$ dependence of $\langle v_2\rangle$ in the 
centrality window and $\eta$ interval as indicated, compared 
with data \cite{Back:2004mh}. The 
solid line indicates result for fluctuating IC, whereas 
the dotted one that for the averaged IC. The curves are  
averages over PHOBOS centrality sub-intervals with  
freeze-out temperatures as indicated.  Fig. taken from \cite{Andrade:2008xh}.} \label{fig_v2_nexspherio2} 
\end{figure}

The first hybrid approach to study initial state fluctuations has been developed by the Brasilian group around 2000 \cite{Andrade:2008xh,Aguiar:2001ac,Andrade:2006yh}. The initial conditions for the 3+1 dimensional ideal smooth particle hydroydnamics calculation is provided by NEXUS, a hadron string approach based on Gribov-Regge theory. A typical initial transverse energy density profile is shown in Fig. \ref{fig_v2_nexspherio1}. The influence of initial state fluctuations on elliptic flow as a function of transverse momentum is shown in Fig. \ref{fig_v2_nexspherio2}.  In this study the elliptic flow at higher transverse momentum is decreased when fluctuations are included. More recently, the same approach has been successfully employed to calculate the transverse momentum dependence of higher Fourier coefficients ($v_1-v_5$) at the highest RHIC energies \cite{Gardim:2011xv}. 

Similarly to the particlization transition on the Cooper-Frye hypersurface there is a crucial transition criterion in the early stages of the collision. In principle, it is straight forward to state that local equilibrium is required to apply a fluid dynamic description. The caveat is that the equilibration process itself is poorly understood. How two nuclei that are crashing into each other at almost the speed of light form an equilibrated quark gluon plasma is one of the big open questions in the field. Therefore, in practice one assumes local equilibration at a certain time and that is the time where the initial $T^{\mu\nu}$ for the fluid dynamic calculation is provided as an input. Another coneptual issue is, that local equilibrium very likely is not achieved everywhere in phase space at the same time. 

\begin{figure}[h]
\includegraphics[width=0.5\textwidth]{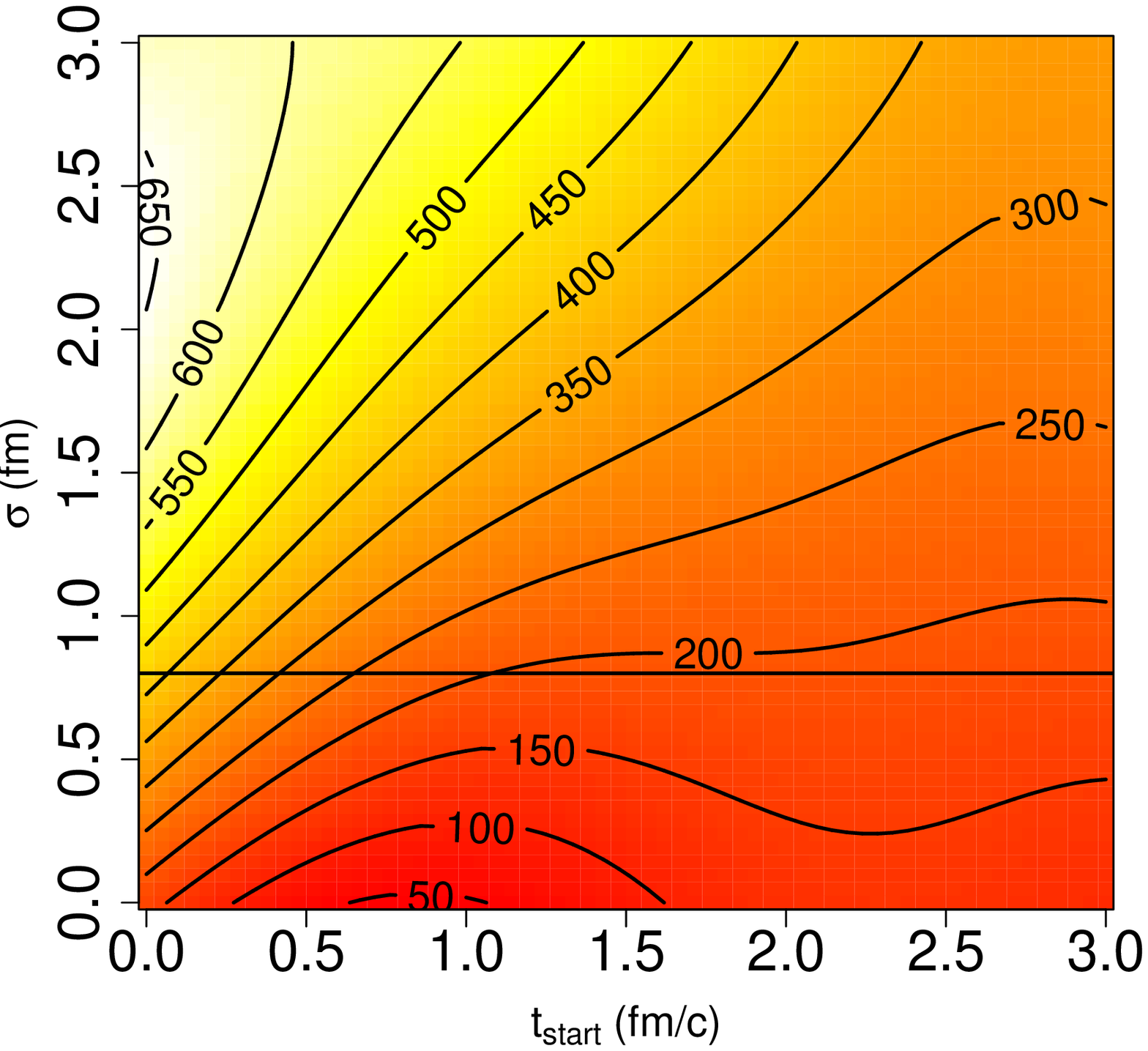}
\includegraphics[width=0.45\textwidth]{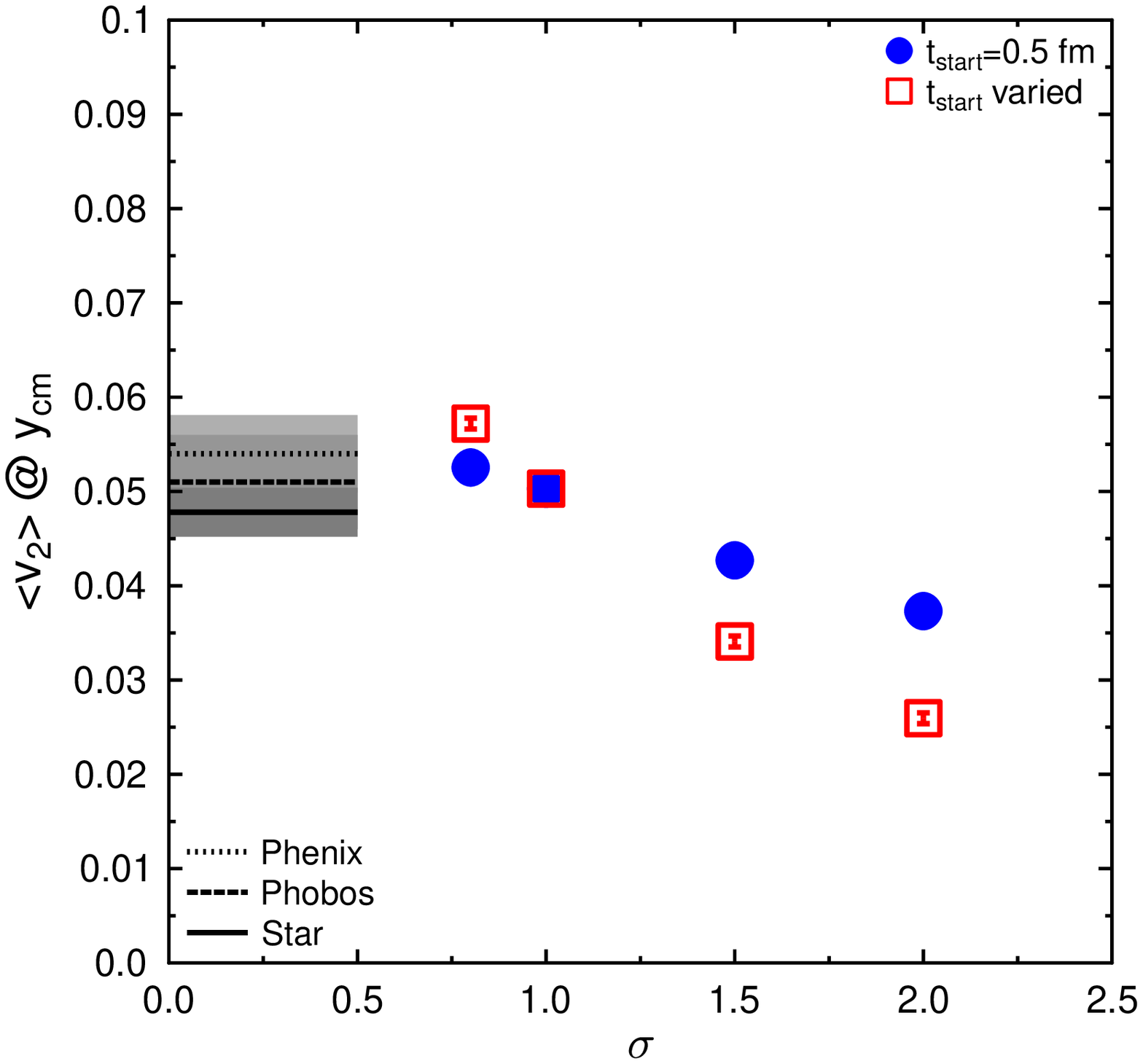}
\caption{Left: Emulated number of pions at midrapidity
($|y|<0.5$) for central ($b<3.4$ fm) Au+Au collisions at  $\sqrt{s_{\rm
NN}}=200$ GeV in the two-dimensional parameter space of $\sigma$ and $t_{\rm
start}$. Right: The averaged value of elliptic flow of
charged particles at midrapidity ($|y|<0.5$) for mid-central ($b=5-9$ fm) Au+Au
collisions at  $\sqrt{s_{\rm NN}}=200$ GeV as a function of $\sigma$ with fixed
(full circles) and varied $t_{\rm start}$ (open squares) compared to
experimental data represented by black lines (the grey shaded regions indicate
the error bars)\cite{Esumi:2002vy,Manly:2002uq,Ray:2002md}. Figs. taken from \cite{Petersen:2010zt}.} \label{fig_ini_params} 
\end{figure}

To demonstrate how the initial conditions are constrained by experimental data, Fig. \ref{fig_ini_params} shows results from the UrQMD hybrid approach for Au+Au collisions at the highest RHIC energy. The initial state is generated by the hadron string transport approach UrQMD and the two free parameters are the starting time $t_{\rm start}$ and the Gaussian kernel width $\sigma$ that determines how large the initial fluctuations are. The left part of the figure shows the number of pions at midrapdity, a measure of the entropy produced in the collision, as a function of these two parameters. Interestingly, it can be observed that later starting times can be compensated by larger Gaussian widths to reach the same final multiplicity. Combined with the right hand side figure where results for integrated charged particle elliptic flow in mid-central collisions are compared to the data, it is clear that starting times around 0.5 fm (and Gaussian width around 1 fm) are necessary for a simultaneous description of both observables.  At this point, all these studies give qualitative insights on sensitivities, but to extract the physics of initial state fluctuations is a much bigger task of high complexity that has recently been reviewed in \cite{Luzum:2013yya}. There is a lot of recent activity to study initial state fluctuations in more detail and there are prospects to develop hybrid approaches based on dynamical Yang-Mills simulations that possibly allow to understand the equilibration process and to match the full energy momentum tensor to viscous fluid dynamics including off-diagonal elements \cite{Gale:2012rq}. 

\begin{figure}[h]
\centering
\includegraphics[width=0.8\textwidth]{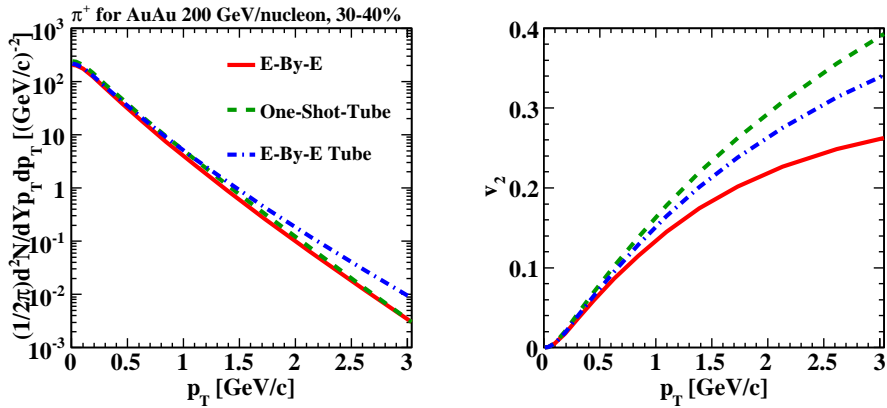}
\caption{The $p_{T}$  spectra (left panel) and elliptic flow (right panel) for charged pion given by hydrodynamic simulations at $30-40\%$ semi-central Au+Au collisions for full AMPT initial conditions (solid lines), event-by-event tube initial conditions (dot-dashed lines) and one-shot-tube initial conditions (dashed lines). Fig. taken from \cite{Pang:2012uw}.} \label{fig_amptic_v2} 
\end{figure}

An additional aspect that can be explored with hybrid approaches where a transport model is employed to generate the initial conditions for the fluid dynamic evolution, is the longitudinal dynamics. Since nowadays, 3+1 dimensional (viscous) hydrodynamic calculations become feasible, there is no  a priori reason to use simplistic parametrizations for the longitudinal initial energy density distribution. Considering all the interesting observations for two-particle correlations and higher flow harmonics that rely on a longitudinal invariance of the fluctuations, it is crucial to study what happens in a more realistic scenario, where the fluctuations are not restricted to the transverse direction. 

Fig. \ref{fig_amptic_v2} shows results from a recent calculation using 3+1 dimensional ideal hydrodynamics based on initial conditions from the AMPT transport approach \cite{Pang:2012uw}. The dashed line indicates the result for transverse momentum spectra and elliptic flow from a smooth initial condition, that has a tube structure in the longitudinal direction. Fluctuations in the transverse plane increase the mean tranverse momentum and decrease elliptic flow (as shown above). The full event-by-event calculation adding also longitudinal fluctuations brings the transverse momentum back down, but decreases elliptic flow even more. Even though the detailed sensitivities have to explored further, this is one example that shows how useful hybrid approaches based on transport models for the initial state are \cite{Petersen:2011fp}. 

\section[Full Hybrids]{Full Hybrid Approaches}
\label{full_hybrid}

Full hybrid approaches that treat the initial and final non-equilibrium dynamics with a microscopic transport approach and emebed a hydrodynamic evolution for the hot and dense stage of the reaction have been developed \cite{Petersen:2008dd,Werner:2010aa}. These approaches allow to treat only the parts of the system where equilibrium can safely be assumed with hydrodynamics, especially with the core-corona option \cite{Steinheimer:2011mp}. EPOS even allows for a combined description of soft and hard physics as will be demonstrated below \cite{Werner:2012xh}. 

\begin{figure}[h]
\centering
\includegraphics[width=0.6\textwidth]{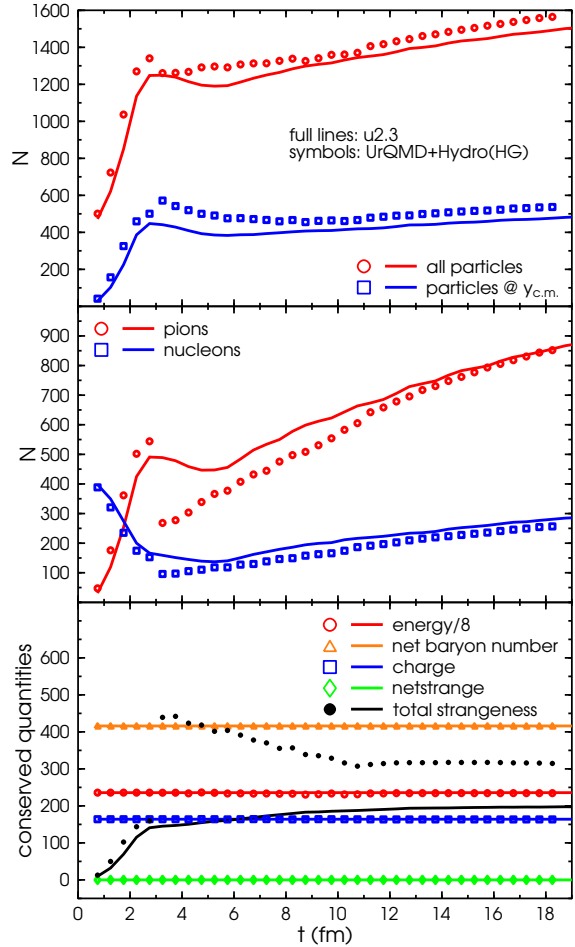}
\caption{Time evolution of the total particle number and the midrapidity ($|y|<0.5$) yield (upper panel), of the total number of pions and nucleons (middle panel) and of the conserved quantities (lower panel) for central ($b=0$ fm) Pb+Pb collisions at $E_{\rm lab}= 40A~$GeV. Results of the hybrid model calculation UrQMD+Hydro (HG) are depicted with symbols, while UrQMD-2.3 results are represented by lines. The total energy of the system (red circles and line) has been divided by eight for visibility reasons. The other conserved quantum numbers are net baryon number (orange triangles and line), the overall charge (blue squares and line) and the strangeness (green diamonds and line). The total strangeness (black dots and line) is given by the sum of s- and $\bar{s}$-quarks. Fig. taken from \cite{Petersen:2008dd}.} \label{fig_time_evol_hybrid} 
\end{figure}

The UrQMD hybrid has been developed in 2008 and first applied to lower beam energies to systematically investigate the differences in the dynamics between a hadron transport and a hybrid approach. Fig. \ref{fig_time_evol_hybrid} shows the time evolution of different quantities, where the full lines show the pure transport calculation and the symbols indicate what happens, if an ideal hydrodynamic evolution is applied for the hot and dense stage with a hadron gas equation of state (HG). First of all, it is interesting to note that the full multiplicities and the yields at midrapidity are very similar in both calculations. In addition all conserved quantities are exactly conserved globally for each event. In contrast the instant thermalization assumption including chemical equilibration of strangeness leads to enhanced strangeness production in the hybrid approach. For the pions and nucleons the thermalization effect is more abrupt in the hybrid calculation, but qualitatively similar in both cases.

\begin{figure}
\centering
\includegraphics[width=0.7\textwidth]{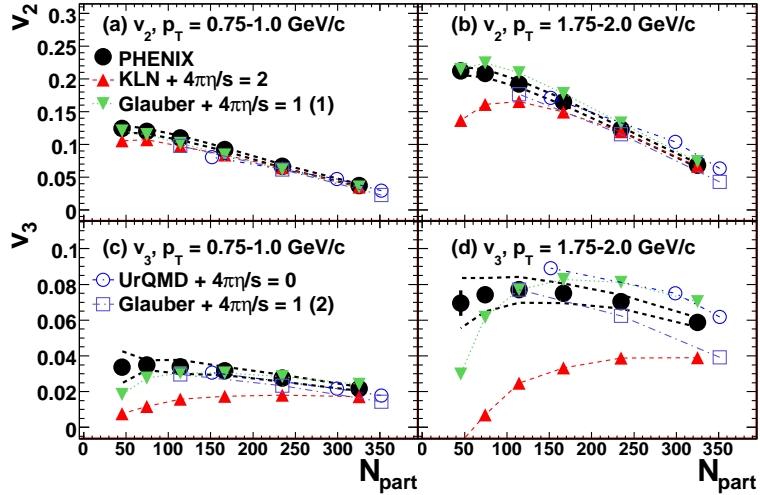}
\caption{Comparison of $v_{n}\{\Psi_n\}$ vs.  $N_{\rm part}$ measurements and 
theoretical predictions (see text): ``MC-KLN + $4\pi\frac{\eta}{s} = 2$" 
and ``Glauber + $4\pi\frac{\eta}{s} = 1$ (1)" \cite{Alver:2010dn}; 
``Glauber + $4\pi\frac{\eta}{s} = 1$ (2)" \cite{Schenke:2010rr}; 
``UrQMD"~\cite{Petersen:2010cw};. The dashed lines (black) around the data 
points indicate the size of the systematic uncertainty.
Fig. taken from \cite{Adare:2011tg}.} \label{fig_phenix_vn} 
\end{figure}

A fully integrated hybrid approach includes the main ingredients that have been identified for the dynamical description of heavy ion collisions: Fluctuating initial conditions, a (nearly) ideal hydrodynamic evolution and hadronic rescattering. Therefore, higher flow harmonics can be calculated in this event by event approach in a straight forward manner. In Fig. \ref{fig_phenix_vn} predictions for triangular flow within the UrQMD hybrid approach \cite{Petersen:2010cw} are compared to experimental data measured by the PHENIX collaboration \cite{Adare:2011tg}. In comparison to other hydrodynamic calculations \cite{Alver:2010dn,Schenke:2010rr} it has been concluded that the integrated elliptic and triangular flow as a function of centrality provide non trivial constraints on the initial conditions and the shear viscosity simultaneously.

The sensitivities of elliptic and triangular flow to the properties of the initial conditions have been explored in more detail in the event  by event hybrid approach for heavy ion collisions at the highest RHIC energy. In \cite{Petersen:2012qc}, it was demonstrated that elliptic flow in mid-central collisions and tranverse mass spectra are insensitive to the granularity in the collision, if one compares single event initial conditions to calculations based on a smooth profile. In addition, it was systematically investigated that triangular flow is directly proportional to the initial granularity, on the average not distorted by final state rescatterings. In contrast, for event shape engineering where the goal is to access initial state properties by selecting certain values for the flow coefficients event by event the hadronic rescattering dynamics needs to be understand in more detail \cite{Petersen:2013vca}.  

\begin{figure}[h]
\includegraphics[width=0.5\textwidth]{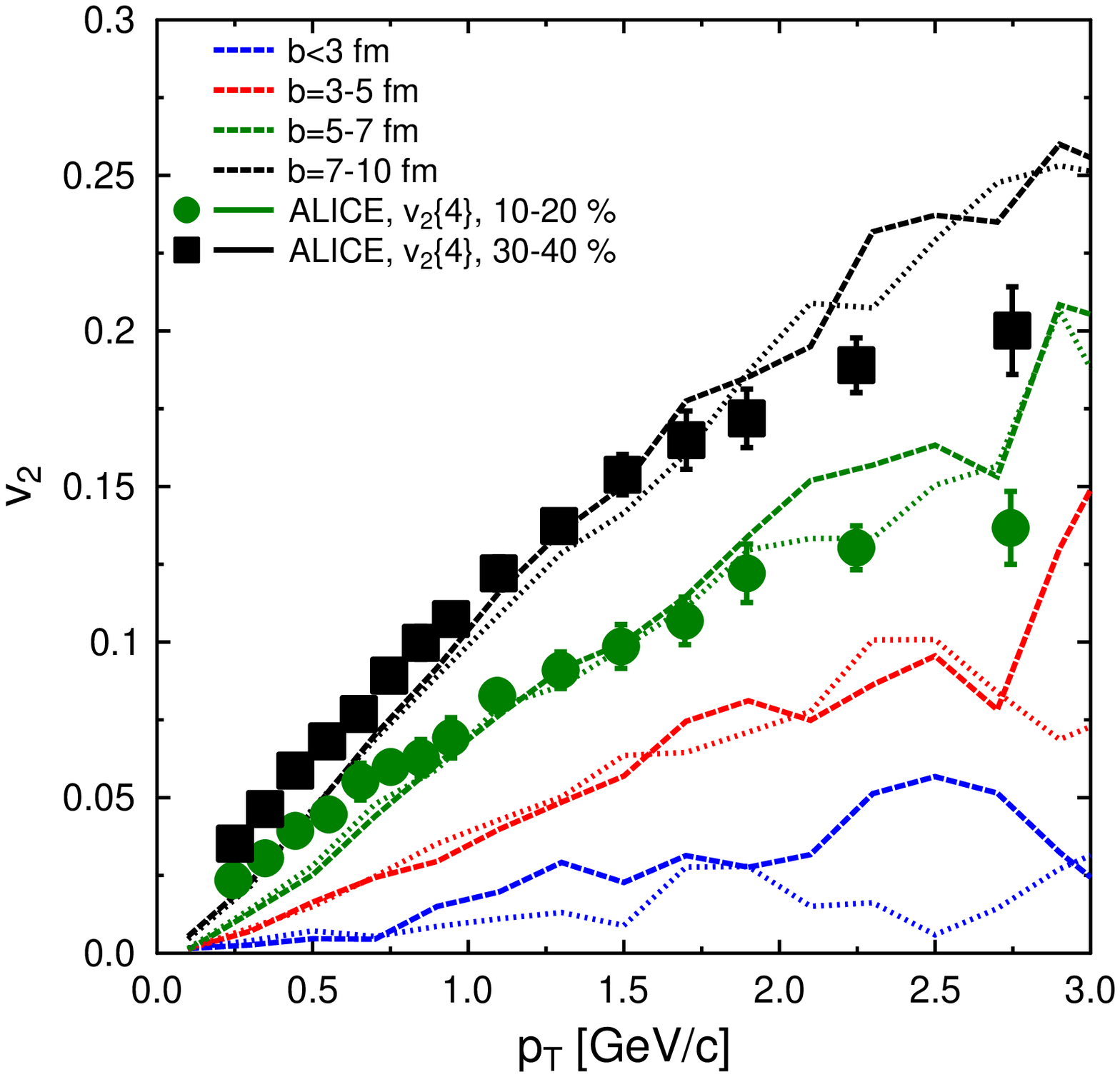}
\includegraphics[width=0.6\textwidth]{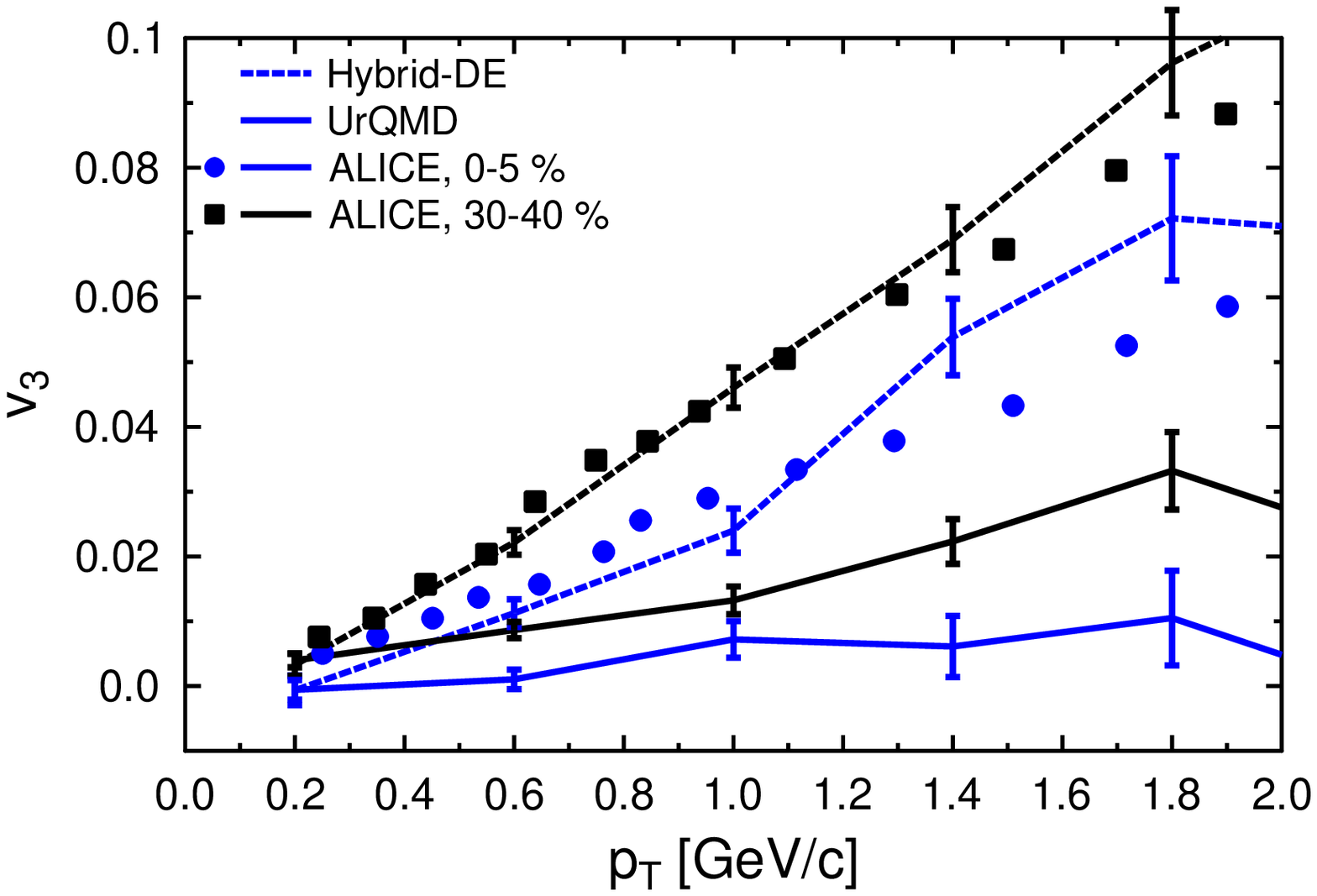}
\caption{Left: Elliptic flow of charged particles as a function of
transverse momentum for
four different centralities calculated in the hybrid
approach with two different equations of state where the dotted line represents the hadron gas (HG) and the dashed line the one with deconfined phase (DE) compared to the experimental
data\cite{Aamodt:2010pa}. The lines are ordered according to increasing centrality from top to bottom. Fig. taken from \cite{Petersen:2011sb}. Right: Triangular flow of charged particles as a function of
transverse momentum calculated in the hybrid approach for four different
centrality classes of Pb+Pb collisions at
$\sqrt{s_{\rm NN}}=2.76$ TeV compared in two centrality bins to the
corresponding results of the UrQMD
transport approach where the lower full line indicates central collisions and the upper full line represents mid-central collisions. The dotted lines are ordered according to increasing centrality from top to bottom compared to ALICE data. } 
\label{fig_lhc_urqmd} 
\end{figure}

The simulation parameters are carefully adjusted to bulk experimental data at high RHIC energies as e.g. in \cite{Petersen:2010zt} to apply the same 3+1 dimensional event by event hybrid approach to calculate elliptic and triangular flow at the LHC \cite{Petersen:2011sb}. The reasonable agreement to measurements from the ALICE collaboration for the transverse momentum dependence at different centralities can be interpreted as a sign that the bulk behaviour in heavy ion reactions at high RHIC and LHC energies can be understood within event by event 3+1 dimensional hybrid approaches. In addition, the full lines in Fig. \ref{fig_lhc_urqmd} (right) depict the result of pure transport which generates too little triangular flow. 

Another full event by event hybrid approach (EPOS) is based on inital conditions from a hadron-string approach and includes a 3+1 dimensional ideal hydrodynamic evolution \cite{Werner:2012xh}. The hadronic rescattering is treated with UrQMD again. This approach has been pushed to explore the limits of applicability of fluid dynamics in high multiplicity p-p \cite{Werner:2010ny} and more recently also ultra high energy p-Pb collisions \cite{Werner:2013ipa}. The main feature of EPOS is that hard processes are treated as well and the interactions of hard particles and soft medium are explored.  

\begin{figure}[h]
\centering
\includegraphics[angle=270,scale=0.4]{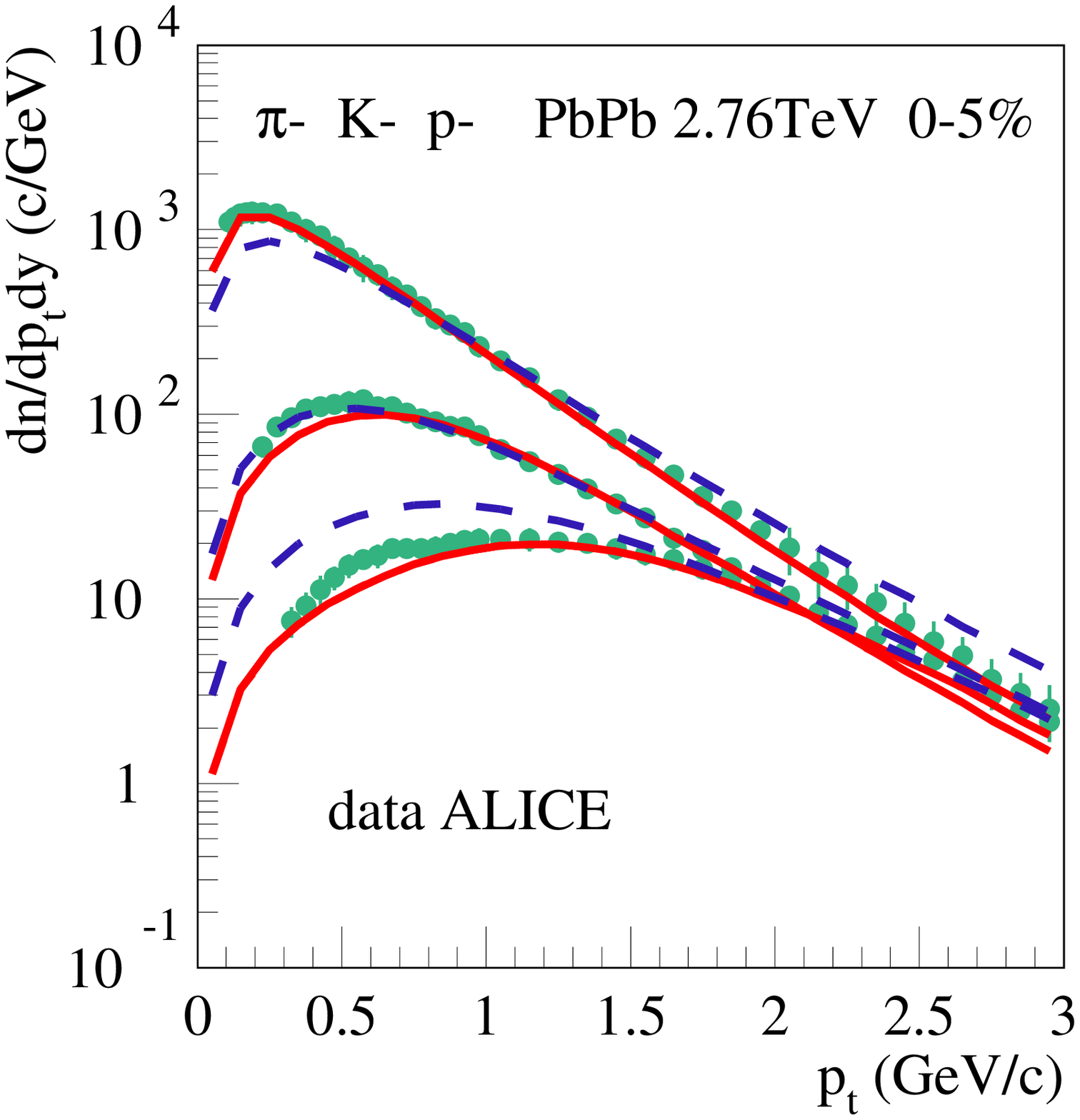}
\includegraphics[angle=270,scale=0.4]{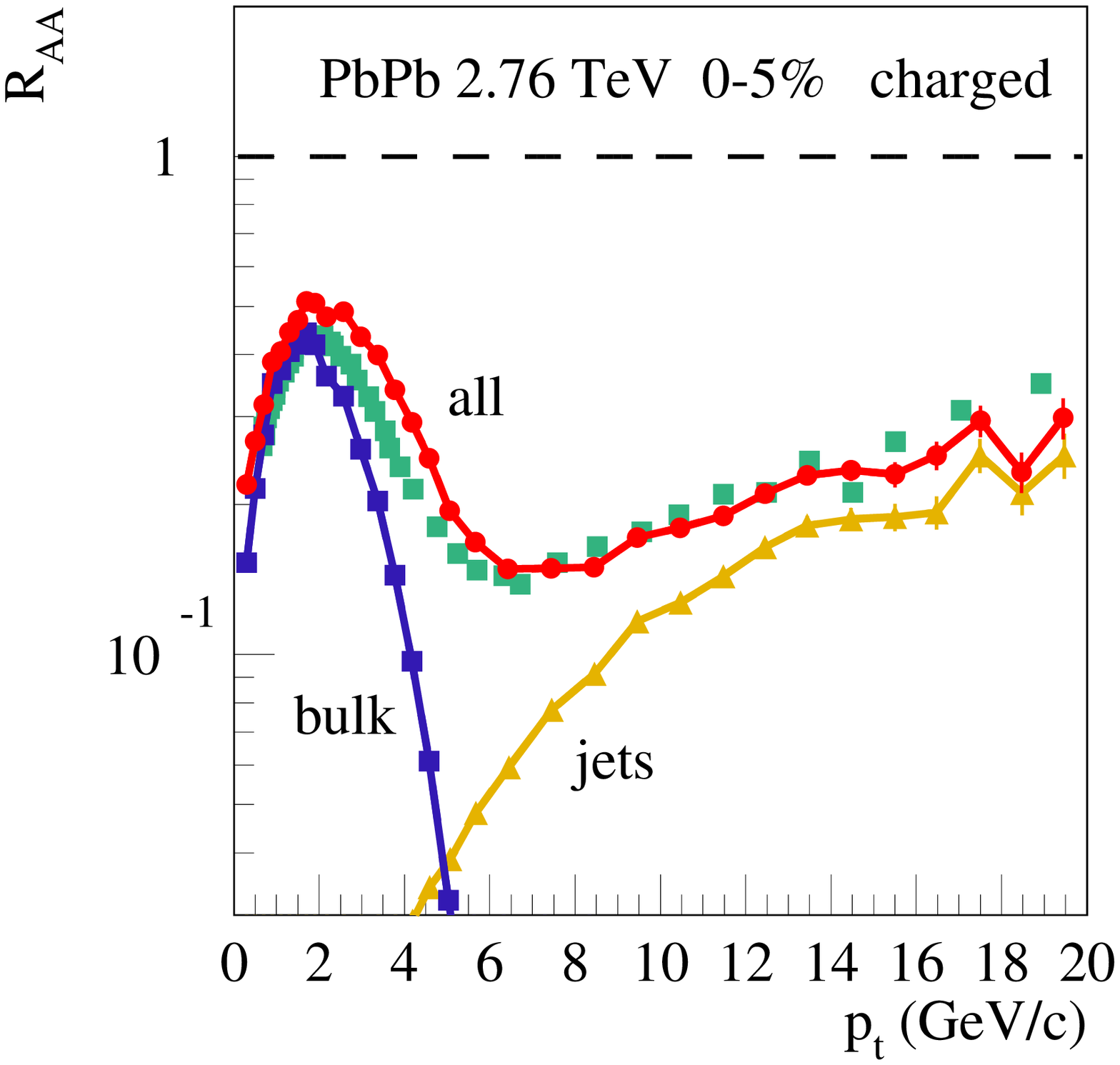}
\caption{Left: Transverse momentum distributions of (from top to bottom)
negative pion, kaons, and protons, in the 0--5\% most central Pb-Pb
collisions at 2.76 TeV. We show the full calculation (solid lines)
and the ones without hadronic cascade (dashed lines), compared to
ALICE data (circles) \cite{Preghenella:2011np}.
Right: The nuclear modification factor in Pb-Pb at 2.76 TeV
vs $p_{t}$: We compare data \cite{Aamodt:2010jd} (squares) with the
full calculation (red line + circles) and its jet contribution (yellow
line + triangles) , as well as the bulk (hydro) contribution of a
calculation without hadronic cascade (blue line + squares). Figs taken from \cite{Werner:2012xh}. }
\label{fig_epos_raapt}
\end{figure}

Fig. \ref{fig_epos_raapt} shows full EPOS hybrid calculations for transverse momentum spectra at LHC. The difference between the dashed and solid lines on the left hand side is attributed to hadronic rescattering that affects different particle species differently, consistent with earlier observations that are mentioned above. On the right hand side the nuclear modification factor $R_{\rm AA}$ is plotted for a large range in transverse momentum, up to 20 GeV/c. Combining soft bulk physics and hard jets as implemented in the EPOS hybrid approach allows for a good description of the data from low to high transverse momentum. The nuclear modification factor is a measure of the suppression of highly energetic partons that travel through the quark gluon plasma. 
 
\begin{figure}
\centering
\includegraphics[angle=270,scale=0.66]{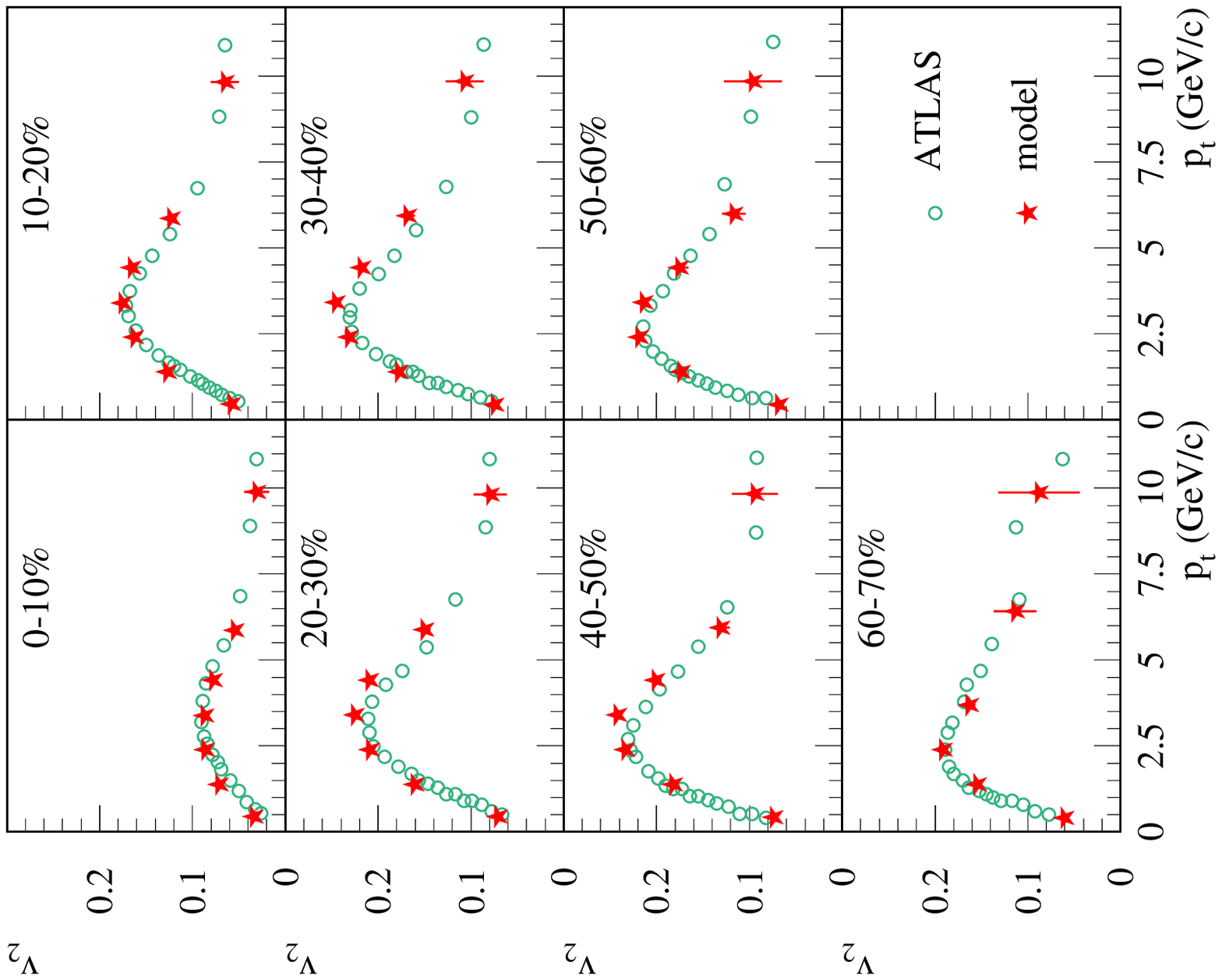}
\caption{$p_T$ dependence of elliptical flow (defined with
respect to the opposite hemisphere sub-event plane) for different
centralities in Pb-Pb collisions at 2.76 TeV. We compare the ATLAS
data \cite{ATLAS:2011ah} (circles) with calculations (red stars).
Fig. taken from \cite{Werner:2012xh}.} \label{fig_epos_v2} 
\end{figure}

In non-central collisions the formed medium has an ellipsoidal anisotropy in the transverse plane, therefore high $p_T$ particles travel different pathlength through the quark gluon plasma depending on their emission angle. This asymmetry is quantified by $v_2$ at high transverse momentum. Fig. \ref{fig_epos_v2} shows the comparison of the Fourier coefficient $v_2$ that measures bulk interaction strength at low $p_T$ and differential energy loss of partons at high pt to EPOS calculations (red stars). The model calculations agree reasonably well with measurements by ATLAS over a large range of centralities and transverse momenta. 

Overall, it seems that full hybrid approaches based on transport for the initial and final off-equilibrium stages and nearly ideal hydrodynamics in the hot and dense stage are very successful for the description of the dynamics of heavy ion collisions at RHIC and LHC energies.

\section[Flow Excitation Function]{Excitation Function of Anisotropic Flow}
\label{flow_exc}

At lower beam energies the 'standard model' for heavy ion reactions, a hybrid transport + hydrodynamics approach, that has been established at the highest RHIC energy and confirmed at LHC faces some challenges. The finite net baryon density needs to be considered in the evolution and an equation of state that extends to finite net baryochemical potential needs to be employed. In addition the non-equilbrium dynamics becomes more relevant at lower beam energies, since the thermalization time increases and the duration of the locally equilibrated quark gluon plasma phase gets shortened.

\begin{figure}
\centering  
\includegraphics[width=0.7\textwidth]{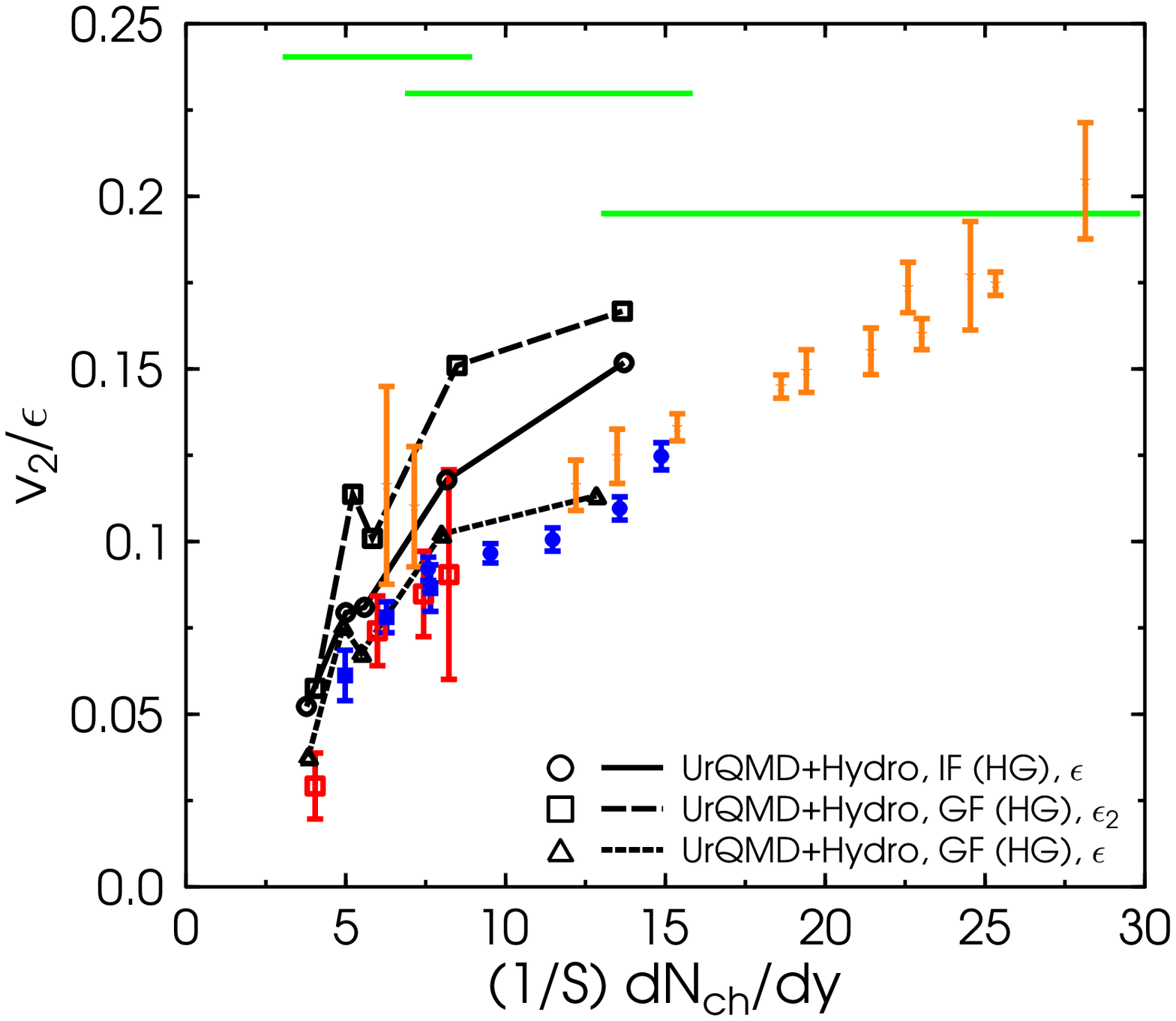}
\caption{$v_2/\epsilon$ as a function of $(1/S)dN_{\rm ch}/dy$ for different energies and centralities for Pb+Pb/Au+Au collisions. The results from mid-central collisions (b=5-9 fm) calculated within the hybrid model with isochronous/gradual freeze-out (IF/GF) (full line with circles and dashed line with triangles respectively) and a hadron gas equation of state (HG) are shown. Furthermore, the hybrid model calculation with GF is divided by a different eccentricity ($\epsilon_{2}$)(dashed line with squares). These curves are compared to data depicted by colored symbols from  different experiments (E877, NA49 and STAR \cite{Alt:2003ab}) for mid-central collisions. The green full lines correspond to the previously calculated hydrodynamic limits \cite{Kolb:2000sd}. Fig. taken from \cite{Petersen:2009vx}. }
\label{fig_urqmd_scaling}      
\end{figure}

Scaling elliptic flow by the corresponding initial state eccentricity allows to compare the medium properties over a large range of beam energies and centralities, since the response function is supposed to be proportional to the charged particle density in the created system measured by $\frac{1}{S} \frac{dN_{\rm ch}}{dy}$. In Fig. \ref{fig_urqmd_scaling} the expected scaling from pure hydrodynamic calculations is shown compared to experimental data. The coincidence of both curves at high RHIC energies was the basis for the claim of the discovery of the perfect liquid. If the early and late stages of the reaction are treated with non-equilibrium hadronic transport the full beam energy dependence of the scaled elliptic flow can be described (as already demonstrated by \cite{Teaney:2000cw}, see above). One issue with this kind of scaling plot is that the eccentricity as well as the overlap area have to be calculated in a theoretical model and cannot be measured. The differences between the black lines illustrate the dependence of the results on the definition of the eccentricity \cite{Petersen:2009vx}.

\begin{figure}
\centering  
\includegraphics[width=0.7\textwidth]{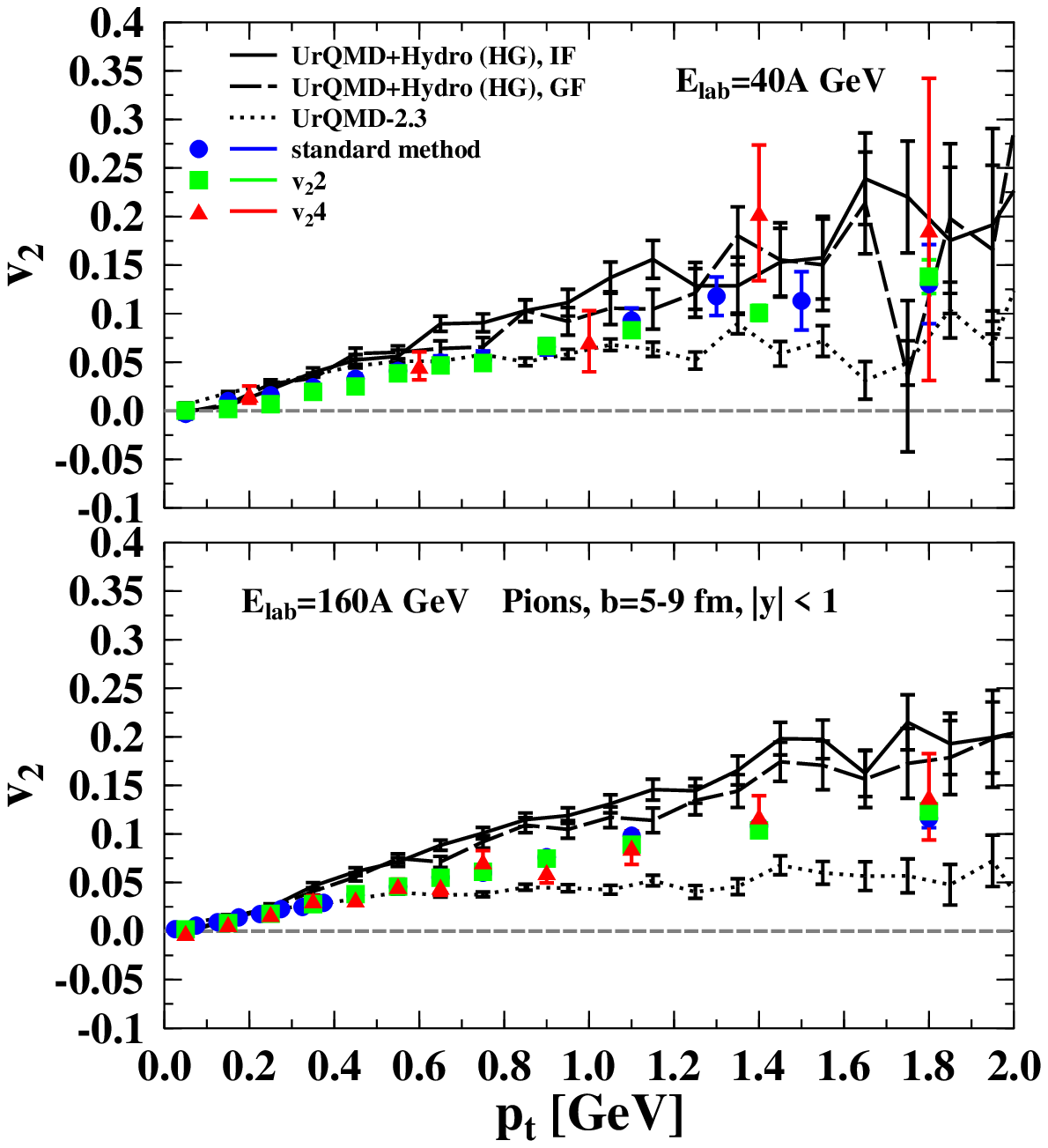}
\caption{Elliptic flow of pions in mid-central (b=5-9) Pb+Pb collisions at $E_{\rm lab}=40A~$GeV and  $E_{\rm lab}=160A~$GeV. The full black line depicts the hybrid model calculation with isochronous freeze-out, the black dashed line the hybrid calculation employing the gradual freeze-out while the pure transport calculation is shown as the black dotted line. The colored symbols display experimental data obtained with different measurement methods by NA49 \cite{Alt:2003ab}. Fig. taken from \cite{Petersen:2009vx}.}
\label{fig_urqmd_v2_ptpi}      
\end{figure}

The differential elliptic flow of pions has been explored within the integrated UrQMD hybrid approach and compared to experimental data from NA49 at the SPS (see Fig. \ref{fig_urqmd_v2_ptpi}). At  $E_{\rm lab}=40A~$GeV, the results are independent of the details of the dynamics, since the pure hadronic transport gives the same result as the hybrid approach with a hadron gas equation of state during the hydrodynamic evolution. In addition, changing the switching criterion between hydrodynamic evolution and hadronic rescattering in the late stages from an isochronous to a gradual (iso-eigentime) transition does not affect the results. Assuming the same degrees of freedom, the transverse momentum dependent elliptic flow becomes sensitive to the hydrodynamic evolution at $E_{\rm lab}=160A~$GeV. At these higher energies, the experimental data requires more interactions during the hot and dense stage than in a pure hadron string transport approach. 

\begin{figure}[h]
\centering  
\includegraphics[width=0.7\textwidth]{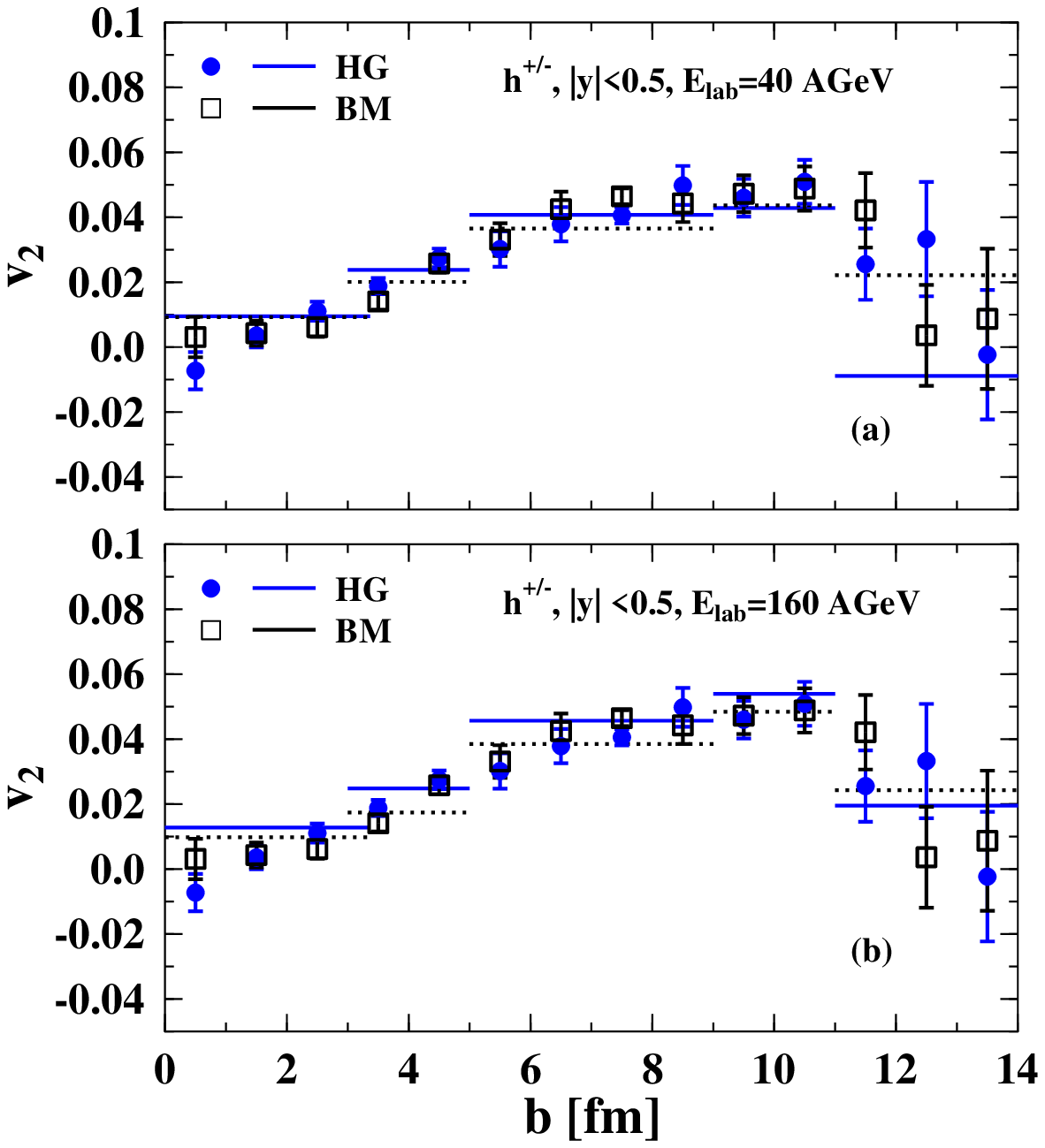}
\caption{Centrality dependence of elliptic flow of charged
particles at midrapidity ($|y|<0.5$) for Pb+Pb collisions at $E_{\rm lab}=40A$
GeV (a) and $E_{\rm lab}=160A$ GeV (b). The horizontal lines indicate the
results for averaged initial conditions using the hadron gas EoS (blue full
line) and the bag model EoS (black dotted line) while the symbols (full circles for
HG-EoS and open squares for BM-EoS) depict the results for the event-by-event
calculation. Fig. taken from \cite{Petersen:2010md}.}
\label{fig_v2_impsps}      
\end{figure}

The same full hybrid approach has been employed to study the centrality dependence of integrated elliptic flow at SPS energies. The advantage of combining initial and final state from UrQMD with a hydrodynamic evolution allows for systematic investigations of different equations of state without retuning initial conditions or freeze-out criterion. In addition the effect of fluctuations in the initial state on elliptic flow can be studied by comparing a single event calculation with one that uses a smooth average over many events as an initial state. Fig. \ref{fig_v2_impsps} shows the difference between a hadron gas equation of state and a bag model equation of state with a first order phase transition.  The final observable anisotropic flow is not influenced significantly (black open squares compared to blue full circles). The horizontal lines indicate the calculation based on averaged initial conditions, whereas the symbols with statistical error bars are final averages of event by event calculations. Even though at that point in time, the theoretical calculation was performed using the reaction plane definition of $v_2$, the main result is still valid: For mid-central collisions where elliptic flow is maximal, the fluctuations do not affect the final elliptic flow value, whereas in central collisions and peripheral collisions the flow is influenced by initial state fluctuations.  

\begin{figure}[h]
\includegraphics[width=0.5\textwidth]{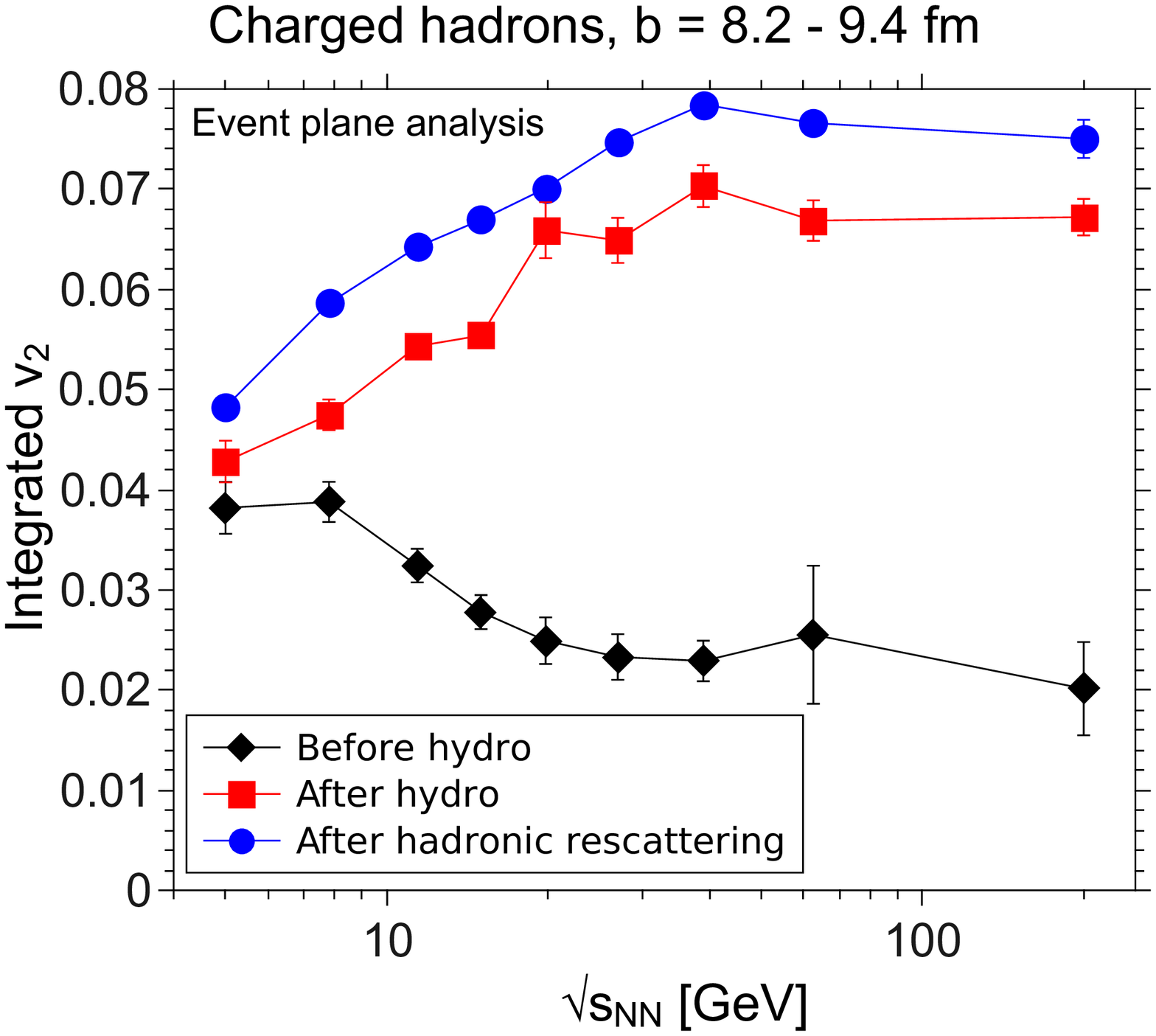}
\includegraphics[width=0.5\textwidth]{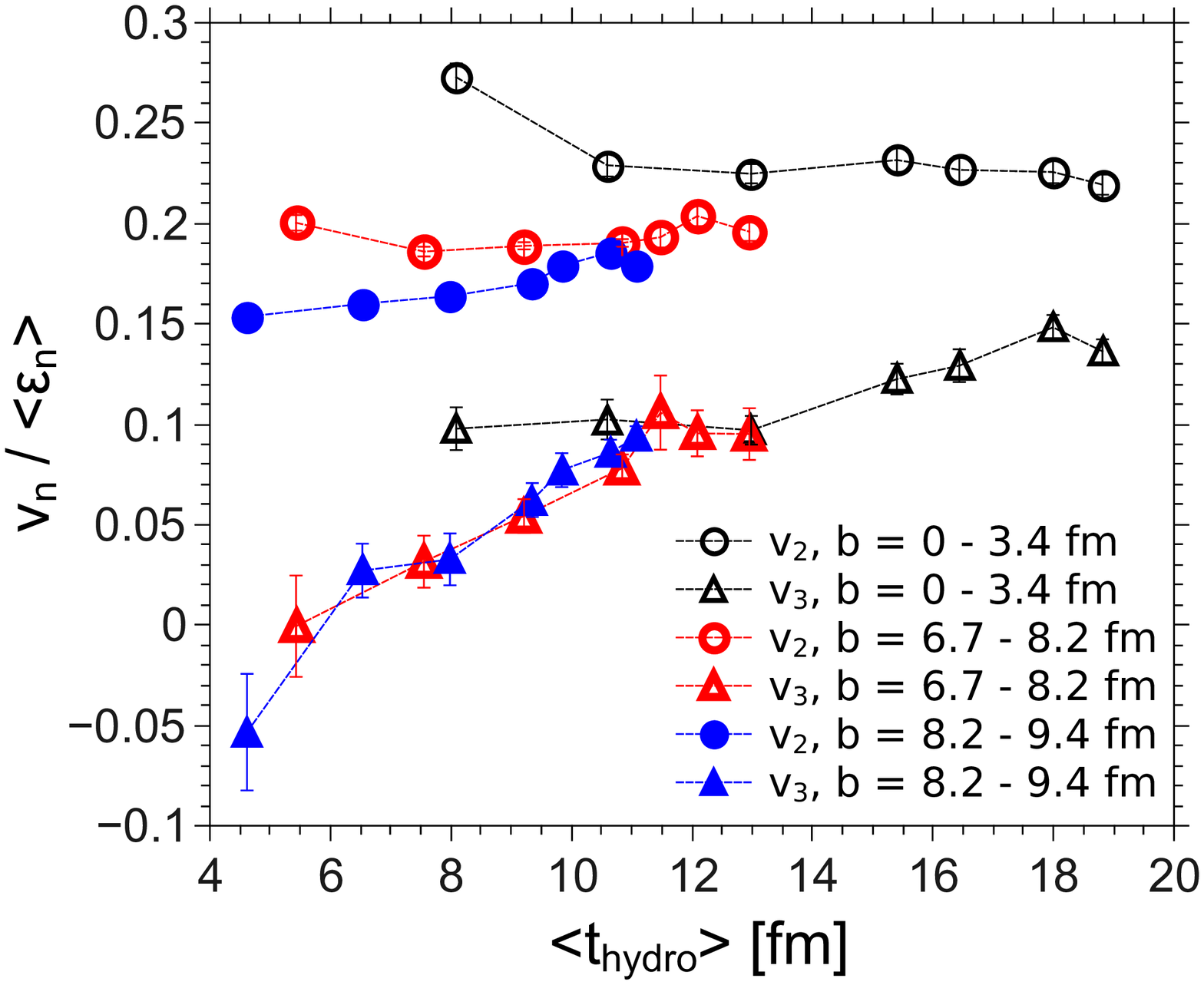}
\caption{Left: Elliptic flow as a function of beam energy in the UrQMD hybrid model. The black diamonds depict the contribution from the early non-equilibrium evolution, the red squares show the result after the hydrodynamic evolution and the blue circles depict the full result including hadronic rescattering. Right: Scaled flow coefficients $v_2/ \langle \epsilon_2 \rangle$ and $v_3/ \langle \epsilon_3 \rangle$ with respect to the average total hydro duration for impact parameter ranges $b = 0-3.4$ fm, $6.7-8.2$ fm and $8.2-9.4$ fm. Figs. taken from \cite{Auvinen:2013sba}.}
\label{fig_besv2} 
\end{figure}

More recently, the UrQMD hybrid approach has been applied to study the excitation function of anisotropic flow. The main difference compared to earlier studies is that a more realistic constant energy density criterion has been employed as a switching criterion for the final stages. In this hybrid approach the initial non-equilibrium evolution lasts longer at lower beam energies, because thermalization is expected to take longer. The geometrical overlap of the two nuclei is defined as the criterion for local equilibrium. Otherwise, there is no beam energy dependence of any parameter for the hydrodynamic evolution and an equation of state at finite net baryon density is employed. In Fig. \ref{fig_besv2} (left) it is shown how much elliptic flow is generated during which stage of the evolution. The black diamonds indicate the decreasing amount of flow that is generated in the early non-equilibrium stage, the red squares show the result at the end of the hydrodynamic evolution and the blue circles denote the final result. The amount of elliptic flow that is generated during the hadronic rescattering phase is rather independent of the energy, whereas the flow that is generated during the hydrodynamic evolution is decreased to almost zero at the lowest energies studied here. This non-trivial interplay of hadronic transport and hydrodynamic evolution might explain the observed energy independence of the differential elliptic flow as a function of transverse momentum.

In addition Fig. \ref{fig_besv2} (right) shows the response of elliptic and triangular flow to the corresponding eccentricity/triangularity as a function of the average duration of the hydrodynamic evolution in heavy ion collisions at different energies and different centralities. For triangular flow a nice scaling can be observed whereas for elliptic flow the deviation from the scaling curves increases at lower beam energies. This result of a hybrid calculation reflects that elliptic flow is more affected by the hadronic evolution than triangular flow which is almost exclusively generated in the strongly coupled fluid dynamic evolution. The newly developed 3 dimensional viscous hybrid approach \cite{Karpenko:2013ksa,Karpenko:2013ama} at finite net baryon densities will allow for more extensive studies of the collective dynamics at lower beam energies.

\section[conclusions]{Conclusions and Outlook}
\label{concl}

Hybrid models based on transport and hydrodynamics are very successful for the dynamical description of heavy ion collisions at high RHIC and LHC energies. Since collective flow is sensitive to the transport properties of the system the radial and anisotropic flow measurements are crucial to support this success. Traditional hybrid approaches rely on parametrized initial conditions and hadronic transport is used to describe the final non-equilibrium stage allowing for the dynamical treatment of succesive chemical and kinetic freeze-out. The multi-strange particles are direct messengers of the transition hypersurface, since the decoupling happens according to the specific hadronic cross sections. The radial flow of protons is increased by 30 \% during the hadronic rescattering stage and the mass splitting of differential elliptic flow requires a proper treatment of hadronic final state interactions. 

A complementary type of hybrid approaches employs transport models for the early non-equilibrium dynamics. The influence of initial state fluctuations on flow observables has been investigated in these models. Elliptic flow in mid-central collisions is not affected by fluctuations but higher odd flow harmonics only exist due to fluctuations in the initial state. Preliminary studies on longitudinal fluctuations indicate that they are alos relavant for a full understanding of flow and correlation observables. 

The last category of hybrid approaches are full hybrid approaches that combine transport for the early and late stages of the reaction with a hydrodynamic evolution for the hot and dense stage. The EPOS hybrid approach shows an impressive agreement to experimental data at LHC energies for a large transverse momentum range. The UrQMD hybrid approach has been applied to lower beam energies. 

Open questions concerning the hybrid description of heavy ion collisions are mainly related to the applicability limits of a nearly ideal fluid dynamic evolution. A full understanding of the non-equilibrium initial evolution including equilibration of the system is still missing. A systematic exploration of switching transition criteria across different beam energies, system sizes and centralities has to be done to understand the freeze-out dynamics and the relevance of the hadronic rescattering in more detail. Also the impact of the hadronic rescattering on higher harmonic flow observables and the treatment of the off-equilibrium distribution function in the Cooper-Frye formula need to be understood. 

The even more fundamental challenges are most relevant for heavy ion collisions at lower beam energies as studied by the recent beam energy scan at RHIC and in the future at FAIR. Here, the finite net baryon density needs to be taken into account in the evolution as well as an equation of state that extends to non-zero baryo chemical potential. The understanding of phase transitions in a finite system with off equilibrium like a heavy ion collision is still in its infancy.

\section*{Acknowledgements}
HP thanks S. A. Bass for helpful discussions an acknowledges funding of a Helmholtz Young Investigator Group VH-NG-822 from the Helmholtz Association and GSI. This work was supported by the Helmholtz International Center for the Facility for Antiproton and Ion Research (HIC for FAIR) within the framework of the Landes-Offensive zur Entwicklung Wissenschaftlich-Oekonomischer Exzellenz (LOEWE) program launched by the State of Hesse.

\end{document}